\documentclass[aps]{revtex4}%
\usepackage{amsfonts}
\usepackage{amsmath}
\usepackage{amssymb}
\usepackage{graphicx}%
\begin{document}
\preprint{ }

\begin{center}
{\LARGE The Approach of Turbulence to the Locally Homogeneous Asymptote as
Studied using Exact Structure-Function Equations}

\textbf{Reginald J. Hill}

Environmental Research Laboratory

National Oceanic and Atmospheric Administration

Boulder CO, 80305-6407, U.S.A.

\strut
\end{center}

\begin{quote}
\textbf{ABSTRACT}. \ An exact equation is obtained that relates the products
of two-point differences of fluid velocity and those differences with the
difference of pressure gradient and other quantities. \ The averages of such
products are structure functions. \ Equations that follow from the
Navier-Stokes equation and incompressibility but with no other approximations
are called \textquotedblleft exact\textquotedblright\ here. \ Exact equations
for structure functions are obtained, as is an exact incompressibility
condition on the second-order velocity structure function. \ Ensemble,
temporal, and spatial averages are all considered because they produce
different statistical equations and because they respectively apply to
theoretical purposes, experiment, and numerical simulation of turbulence;
those applications are addressed herein. \ The midpoint and the difference of
the two points at which the hydrodynamic quantities are obtained are
$\mathbf{X}$ and $\mathbf{r}$; $t$ is time. \ The equations are organized in a
revealing way by use of $\mathbf{X}$, $\mathbf{r}$, $t$ as independent
variables. \ Dependences on $\mathbf{X}$ and on the orientation of
$\mathbf{r}$ and on $t$ fade as the asymptotic statistical states of local
homogeneity, local isotropy, and local stationarity, respectively, are
approached. \ The exact equations are thus applicable to study of the approach
toward those asymptotic states. \ Exact equations obtained by averaging over a
sphere in $\mathbf{r}$-space have a particularly simple form. \ The case of a
simulation that has periodic boundary conditions leads to particularly simple
equations. \ A new definition of local homogeneity is contrasted with previous
definitions. \ The approach toward the asymptotic state of local homogeneity
is studied by using scale analysis to determine the required approximations
and the approximate equations pertaining to experiments and simulations of the
small-scale structure of high-Reynolds-number turbulence, but without invoking
local isotropy. \ Those equations differ from equations for homogeneous
turbulence. \ The traces of both exact and approximate equations have
particularly simple forms; in particular, the energy dissipation rate appears
in the exact trace equation even without averaging, whereas in previous
formulations the energy dissipation rate appears after averaging and use of
local isotropy. \ The trace mitigates the effect of anisotropy in the
equations, thereby revealing that the trace of the third-order structure
function is expected to be superior for quantifying asymptotic scaling
laws.\bigskip
\end{quote}

\begin{center}
\bigskip\textbf{1. \ INTRODUCTION}
\end{center}

$\qquad$The dynamic theory of the local structure of turbulence is so named by
Monin and Yaglom (1975) (their Sec. 22) to mean the derivation and
investigation of equations for structure functions by use of the Navier-Stokes
equation. \ The structure functions are averages of differences of basic
hydrodynamic quantities such as velocity and pressure gradient. \ Monin and
Yaglom (1975) pointed out that the dynamic theory gives important
relationships between structure functions, and that these relationships
provide important extensions of predictions based on dimensional analysis and
flow similarity. \ The dynamic theory is the basis for Kolmogorov's (1941a)
famous equation that relates second-order and third-order velocity structure
functions, and is of fundamental importance in the theory of locally
homogeneous and locally isotropic turbulence (Monin and Yaglom, 1975;
Batchelor, 1947; Monin, 1959; Frisch, 1995). \ The dynamic theory does not
uniquely determine the structure functions; this is known as the closure
problem (Monin and Yaglom, 1975). \ Experimental data have been used to
evaluate the balance of Kolmogorov's equation and generalizations of it
(Antonia, Chambers, and Browne, 1983; Chambers and Antonia, 1984; Lindborg,
1999; Danaila \textit{et al.}, 1999 a,b; Antonia \textit{et al.}, 2000).
\ This report supports such experimental work, as well as more precise use of
direct numerical simulation (DNS) by giving correct and complete equations to
be used in such evaluations.

\qquad We derive exact equations for structure functions by use of
differential operator identities. \ By ``exact'' we mean that the equations
follow from the Navier-Stokes equation and\ the incompressibility condition
with no additional approximations. \ This meaning is emphasized because
turbulence researchers consistently use ``exact'' when they mean asymptotic.
Exact equations satisfy the perceived need by Yaglom (1998) for careful
derivation of dynamic-theory equations and the perceived value placed by
Sreenivasan and Antonia (1997) on aspects of turbulence that can be understood
precisely. \ In Sec. 2, the equations for products of differences is developed
to the greatest extent possible before any\ average is performed. \ This
mathematical method is similar to that used in the theory of wave propagation
in random media where the equations for wave-field products are thoroughly
developed before an average is performed (see Rytov \textit{et al.}, 1989).
\ Our study is limited to two spatial points and a single time and to\ the
lowest-order equation of the dynamic theory; that equation includes second-
and third-order velocity structure functions. \ On the other hand, the
mathematical method used is of wider applicability; it is not limited to just
two points, a single time, the lowest-order equation, or to the Navier-Stokes
equation alone. \ For instance, the method could be used to derive an equation
involving a three-point structure function for a scalar quantity having its
continuity equation coupled to the velocity field. \ Such an equation would be
useful for interpreting the observed (Mydlarski and Warhaft, 1998;
Sreenivasan, 1991) local anisotropy of scalar fields in the presence of a mean
gradient of the scalar. \ 

\qquad The exact equations retain all of the dependence of the structure
functions on $\mathbf{r,}$ $\mathbf{X}$, and time $t$, where $\mathbf{r}$ is
the vector spacing between two points at which the measurements are obtained
within the turbulent flow and $\mathbf{X}$ is the midpoint position of these
points. Previous methods (Batchelor, 1956; Lindborg, 1996; Hill, 1997) of
deriving dynamic-theory equations neglected the dependence of statistics on
$\mathbf{X}$, and thereby limited the equations to the cases of homogeneous
and locally homogeneous turbulence. \ To also study the approach toward local
homogeneity, equations are needed that retain $\mathbf{X}$. \ Here, attention
is given to the conditions that must be fulfilled for the $\mathbf{X}%
$-dependence to be neglected. \ Previously (Hill, 1997), the approach toward
local isotropy was examined, although exact equations were not then available.
Consequently, the approach toward local isotropy is not considered here.

\qquad A scale analysis is performed to quantify terms that are to be
neglected on the basis that $\left\vert \mathbf{r}\right\vert $ is much less
than a length scale that will be called the outer scale, and to deduce all
other required approximations. \ Such scale analysis is presented in detail in
Sec. 7.4. \ Our analysis determines approximations that quantify the degree to
which the small-scale structure of turbulence depends on its large-scale
structure; such analysis was called for by Yaglom (1998). \ Our analysis sets
the stage for DNS and experimental studies of the approximations.

\qquad The equations derived in Sections 2 and 3 are exact for every flow,
whether laminar or turbulent.\ \ For example, the equations apply exactly to
the edge of a jet, to a boundary layer, as well as to those experimental
situations such as grid-generated wind-tunnel turbulence, for which local
homogeneity is expected to be most accurate. \ The equations apply provided
there are no forces on the fluid at the points of measurement. \ Forces can be
applied near the point of measurement; for instance, the equations are exact
for hot-wire anemometer supports just downstream of the measurement points.
\ The equations apply for turbulence generated at places other than the points
of measurement; examples are grid-generated turbulence measured downstream of
the grid, and turbulence generated by rotating blades (Zocchi \textit{et al.},
1994). \ The case of statistically homogeneous forces distributed throughout
the fluid has been considered for the asymptotic case of isotropic turbulence
by Novikov (1965) (see also Frisch, 1995). \ The case of forces at the points
of measurement is considered in Appendix A.

\qquad The ensemble average is considered first (Sec. 3.1). \ It has the
advantage for theoretical studies that temporal and spatial changes can be
considered because the ensemble average does not eliminate dependence on
$\mathbf{X}$\ or $t$. \ The temporal average is typically used with
experimental data, and the spatial average is typically used for data from
DNS. \ For this reason, exact equations for both temporal averaging (Sec. 3.2)
and spatial averaging (Sec. 3.3) are also obtained. \ The connection between
the derivations presented here and any experiment or DNS is important because
the equations relate several statistics and therefore are most revealing when
data are substituted into them. \ A recently developed experimental method (Su
and Dahm, 1996) has the potential to thoroughly evaluate terms in the
equations derived here.\ \ As shown in Sec. 3.4, the exact equations have a
particularly simple form for the case of DNS with periodic boundary conditions.

\qquad The equations can be evaluated with experimental or DNS data to
determine the most significant terms in the equations for a given flow and
thereby determine the effects that cause deviations from asymptotic laws.
\ The ongoing interest in turbulence intermittency includes accurate
evaluation of inertial-range exponents of structure functions, for which
purpose precise definition of an observed inertial range is needed. \ The
third-order structure function can serve this purpose because it has a
well-known inertial-range power law and the 4/5 coefficient (Kolmogorov's
(1941a) 4/5 law) in the asymptotic limit of accurate local homogeneity and
local isotropy. \ Deviations from the 4/5 coefficient are observed in
experiments (Anselmet \textit{et al.}, 1984; Mydlarski and Warhaft, 1996,
1998; Lindborg, 1999); this casts doubt on the precision with which measured
exponents apply to the intermittency phenomenon. \ The equations derived here,
when evaluated with data, can reveal the effects contributing to the deviation
from Kolmogorov's 4/5 law. \ The usefulness of such evaluations is shown by
Lindborg (1999); Danaila \textit{et al.} (1999 a,b); and Antonia \textit{et
al.} (2000). \ They generalize Kolmogorov's equation by the addition of a term
describing streamwise inhomogeneity. \ To obtain this term from the present
exact analysis, it is necessary to perform the Reynolds decomposition. \ The
present analysis has the advantage that it reveals all terms that describe
inhomogeneity. \ This is discussed in detail in Sec. 7\textbf{.} \ The
equations derived here are obtained in the Eulerian framework, which is most
useful for experimental evaluation.

\qquad Particular attention is given to the typical experimental case that is
used to investigate universality of turbulence statistics at small scales and
large Reynolds numbers. \ We derive the simplification of the exact equation
that applies approximately to such experiments. \ Experimental data typically
have the mean velocity subtracted before structure functions are calculated
from the velocity fluctuation. \ For this reason, we derive the approximate
equation obeyed by structure functions calculated from velocity fluctuations.
\ The Reynolds decomposition (Sec. 5) is essential for this purpose. \ The
derivation is necessarily long in Sec. 7.4, but in this case, the journey is
more significant than the destination because all required approximations are
determined en route. \ Local homogeneity is the most important of the
approximations. \ A necessary condition for local homogeneity is given in Sec.
7.3; it is not a sufficient condition.

\qquad The trace of the exact equation has a particularly simple form. \ When
averaged over a sphere in $\mathbf{r}$-space, and when the advective and
time-derivative terms are neglected, this equation has the same form as
Kolmogorov's (1941a) equation (Sec. 4.3). \ This is true\ despite the fact
that the $\mathbf{r}$-space sphere-averaged\ equation is valid even for
extreme violations of local isotropy.

\bigskip

\begin{center}
\textbf{1.1 \ Contrasting Definitions of Local Homogeneity\bigskip}
\end{center}

\qquad Local homogeneity has been given various definitions by different
authors. \ Kolmogorov (1941b) introduced a space-time domain that is small
compared to L and T=(L/U), where L and U are \textquotedblleft typical length
and velocity for the flow in the whole.\textquotedblright\ \ Kolmogorov
considers the two-point differences of the velocities at spatial points in the
domain; one point is common to all the differences. \ Kolmogorov (1941b)
defines local homogeneity as follows: \ the joint probability distribution of
the velocity differences is independent of the one common spatial point, and
of the velocity at the one common point, and of time. \ Data of Praskovsky et
al. (1993), Sreenivasan \& Stolovitzky (1996), and Sreenivasan \& Dhruva
(1998) contradict the statistical independence of velocity difference and the
velocity at either end point, as well as contradict the statistical
independence of velocity difference and the velocity at the midpoint. \ The
exception is isotropic turbulence (Sreenivasan \& Dhruva, 1998)\ for which
case local homogeneity is assured. \ An alternative possibility that is
particularly relevant here is that the two-point velocity sum, $u_{n}%
+u_{n}^{\prime}$ might be statistically independent of velocity difference,
but statements by Sreenivasan \& Stolovitzky (1996) and Sreenivasan \& Dhruva
(1998)\ contradict that statistical independence as well; publication of
supporting data would be useful. \ Kolmogorov's definition should not be used
because experimental data contradict that statistical independence (Praskovsky
\textit{et al.}, 1993; Sreenivasan and Stolovitzky, 1996; Hill and Wilczak,
2001), as do theoretical considerations (Hill and Wilczak, 2001).

\qquad Monin and Yaglom (1975) define local homogeneity to mean that the joint
probability distribution of the two-spatial-point velocity differences is
unaffected by any translation of the spatial points. \ They do not impose a
restriction on the translations to a spatial domain. \ It follows (Monin and
Yaglom, 1975) that statistics composed entirely of the differences obey the
same relationships that they do for homogeneous turbulence (namely, they are
independent of where they are measured), and that the mean velocity depends
linearly on position. \ In practice, statistics of differences and of
derivatives do depend on where they are measured except in the ideal case of
homogeneous turbulence. \ Frisch (1995) gives a definition that is equivalent
to that of Monin and Yaglom (1975), except that the translations are
restricted to a domain the size of the spatial scale characteristic of the
production of turbulent energy (which he calls the integral scale).
\ Two-point structure-function equations of all orders contain a statistic
that is the product of not only factors of the difference of the two
velocities but also one factor of the sum of the two velocities, i.e.,
$u_{n}+u_{n}^{\prime}$ (Hill, 2001). \ Because the definitions of local
homogeneity by Monin and Yaglom (1975) and Frisch (1995) involve only the
joint probability distribution of two-point differences, it follows that those
definitions are not sufficient to simplify structure-function equations to the
same level of simplification as does homogeneity.

\qquad The calculus of homogeneity by Batchelor (1956) is the commutation of
spatial derivatives from within an average to outside the average where they
become derivatives with respect to $\mathbf{r}$, and vice versa. \ The
calculus of local homogeneity by Hill (1997) is a generalization of
Batchelor's calculus; specifically, local homogeneity was implemented by
neglecting the derivative with respect to $\mathbf{X}$ relative to the
derivative with respect to $\mathbf{r}$\ when spatial derivatives were
commuted with the averaging operation. \ That implementation is restricted to
statistics that contain at least one difference or derivative of basic
hydrodynamic quantities (such as velocity, pressure, temperature, etc.).
\ This calculus differs from the aforementioned definitions of local
homogeneity in that no translational invariance is required other than for the
infinitesimal displacement in $\mathbf{X}$ implied by the derivative
operation. \ In Appendix C, examples are given that show how this calculus
produces the predictions of homogeneity for the homogeneous case. \ To
simplify the structure-function equations, Hill (1997, 2001) found that it was
necessary to apply that calculus to statistics of products containing not only
at least one difference but also quantities that were not differences.

\qquad Consider grid-generated turbulence in a wind tunnel operated with
constant mean velocity. \ For anemometers fixed relative to the position of
the grid, the turbulence is stationary and streamwise inhomogeneous. \ For
simplicity, ignore the cross-stream inhomogeneity. \ For anemometers moving
relative to the grid in a direction parallel to the streamwise direction, the
turbulence is both streamwise inhomogeneous and nonstationary. \ It is
nonstationary because of downstream decay of the turbulence intensity. \ That
example raises the question as to whether or not local stationarity and local
homogeneity should be combined into a single definition that is independent of
the motion of the coordinate system. \ In this author's opinion such a
combined definition is neither desirable nor practical. \ Thus, local
homogeneity (or local stationarity) must be considered in a given coordinate system.

\bigskip

\begin{center}
\textbf{2. \ EXACT TWO-POINT EQUATIONS\bigskip}
\end{center}

\qquad Exact equations are given here that relate two-point quantities and
that are obtained from the Navier-Stokes equations and incompressibility.
\ The two spatial points are denoted $\mathbf{x}$ and $\mathbf{x}^{\prime}%
$;\ they are independent variables: they have no relative motion; e.g.,
anemometers at $\mathbf{x}$ and $\mathbf{x}^{\prime}$ are fixed relative to
one another. \ To be concise, velocities are denoted $u_{i}=u_{i}%
(\mathbf{x},t)$, $u_{i}^{\prime}=u_{i}(\mathbf{x}^{\prime},t)$, and the same
notation is used for other quantities. \ $p(\mathbf{x},t)$ is the pressure
divided by the density (density is constant), $\nu$ is kinematic viscosity,
and $\partial$ denotes partial differentiation with respect to its subscript
variable. \ Summation is implied by repeated\ Roman indices; e.g.,
$\partial_{x_{n}}\partial_{x_{n}} $ is the Laplacian operator. \ For brevity,
define:
\begin{align}
d_{ij}  &  \equiv\left(  u_{i}-u_{i}^{\prime}\right)  \left(  u_{j}%
-u_{j}^{\prime}\right)  ;\\
d_{ijn}  &  \equiv\left(  u_{i}-u_{i}^{\prime}\right)  \left(  u_{j}%
-u_{j}^{\prime}\right)  \left(  u_{n}-u_{n}^{\prime}\right)  ;\\
\tau_{ij}  &  \equiv\left(  \partial_{x_{i}}p-\partial_{x_{i}^{\prime}%
}p^{\prime}\right)  \left(  u_{j}-u_{j}^{\prime}\right)  +\left(
\partial_{x_{j}}p-\partial_{x_{j}^{\prime}}p^{\prime}\right)  \left(
u_{i}-u_{i}^{\prime}\right)  ;\label{tau}\\
e_{ij}  &  \equiv\left(  \partial_{x_{n}}u_{i}\right)  \left(  \partial
_{x_{n}}u_{j}\right)  +\left(  \partial_{x_{n}^{\prime}}u_{i}^{\prime}\right)
\left(  \partial_{x_{n}^{\prime}}u_{j}^{\prime}\right)  ;\label{eij}\\
\digamma_{ijn}  &  \equiv\left(  u_{i}-u_{i}^{\prime}\right)  \left(
u_{j}-u_{j}^{\prime}\right)  \frac{u_{n}+u_{n}^{\prime}}{2}. \label{effunav}%
\end{align}

We change independent variables from $\mathbf{x}$ and $\mathbf{x}^{\prime}$ to
the sum and difference independent variables:
\begin{equation}
\mathbf{X}\equiv\left(  \mathbf{x}+\mathbf{x}^{\prime}\right)  /2\text{ \ and
\ }\mathbf{r}\equiv\mathbf{x}-\mathbf{x}^{\prime}\text{, \ \ \ \ \ and define
}r\equiv\left\vert \mathbf{r}\right\vert .\label{change}%
\end{equation}
The derivatives $\partial_{X_{i}}$ and $\partial_{r_{i}}$ are related to
$\partial_{x_{i}}$ and $\ \partial_{x_{i}^{\prime}}$\ by$\ \ \ \ \ $%
\begin{equation}
\partial_{x_{i}}=\partial_{r_{i}}+\frac{1}{2}\partial_{X_{i}}\text{ ,
}\ \partial_{x_{i}^{\prime}}=-\partial_{r_{i}}+\frac{1}{2}\partial_{X_{i}%
}\text{ \ , }\partial_{X_{i}}=\partial_{x_{i}}+\partial_{x_{i}^{\prime}}\text{
\ , }\partial_{r_{i}}=\frac{1}{2}\left(  \partial_{x_{i}}-\partial
_{x_{i}^{\prime}}\right)  \text{.}\label{derivs}%
\end{equation}
It is essential to hold fixed the correct variables for each of the above
partial derivative operations. \ The partial derivative $\partial_{x_{i}}$ is
obtained with the following variables held fixed: $x_{j}$, for $j\neq i$, and
$\mathbf{x}^{\prime}$ and $t$. \ Likewise for $\partial_{x_{i}^{\prime}}$,
$x_{j}^{\prime}$, for $j\neq i$, and $\mathbf{x}$ and $t$ are held fixed.
\ For $\partial_{X_{i}}$, $X_{j}$, for $j\neq i$, and $\mathbf{r}$ and $t$ are
held fixed. \ For $\partial_{r_{i}}$, $r_{j}$, for $j\neq i$, and $\mathbf{X}$
and $t$ are held fixed. \ For any functions $f(\mathbf{x},t)$ and
$g(\mathbf{x}^{\prime},t)$, (\ref{derivs}) gives
\begin{equation}
\partial_{r_{i}}\left[  f(\mathbf{x},t)\pm g(\mathbf{x}^{\prime},t)\right]
=\partial_{X_{i}}\left[  f(\mathbf{x},t)\mp g(\mathbf{x}^{\prime},t)\right]
/2.\label{identderivs}%
\end{equation}
For example, $\partial_{r_{i}}\left(  u_{j}-u_{j}^{\prime}\right)
=\partial_{X_{i}}\left(  u_{j}+u_{j}^{\prime}\right)  /2$ , and \ $\partial
_{r_{i}}\left(  u_{j}+u_{j}^{\prime}\right)  =\partial_{X_{i}}\left(
u_{j}-u_{j}^{\prime}\right)  /2$.

\qquad Now, $\tau_{ij}$ and the trace of (\ref{tau}) and (\ref{eij}) (i.e.,
$\tau_{ii}$ and $e_{ii}$) can be expressed differently. \ Use of
(\ref{derivs}) in (\ref{tau})\ as well as in $e_{ii}$ and rearranging terms gives%

\begin{align}
\tau_{ij}  &  =-2\left(  p-p^{\prime}\right)  \left(  s_{ij}-s_{ij}^{\prime
}\right)  +\partial_{X_{i}}\left[  \left(  p-p^{\prime}\right)  \left(
u_{j}-u_{j}^{\prime}\right)  \right]  +\partial_{Xj}\left[  \left(
p-p^{\prime}\right)  \left(  u_{i}-u_{i}^{\prime}\right)  \right]
\text{,}\label{tau2}\\
2\nu e_{ii}  &  =2\left(  \varepsilon+\varepsilon^{\prime}\right)
+2\nu\partial_{X_{n}}\partial_{X_{n}}\left(  p+p^{\prime}\right)  \text{,}
\label{eii}%
\end{align}

\begin{equation}
\text{where \ \ \ \ \ \ \ \ \ \ \ \ \ \ }s_{ij}\equiv\left(  \partial_{x_{i}%
}u_{j}+\partial_{x_{j}}u_{i}\right)  /2\text{ , and \ \ }\varepsilon\equiv2\nu
s_{ij}s_{ij}\text{\ ;\ \ \ \ \ \ \ \ \ \ \ \ \ \ \ \ \ \ \ } \label{strain}%
\end{equation}
to obtain (\ref{eii}) we used Poisson's equation $\partial_{x_{n}}%
\partial_{x_{n}}p=-\partial_{x_{i}}u_{j}\partial_{x_{j}}u_{i}$.
\ Incompressibility requires that the trace of $s_{ij}$ vanishes; thus, the
trace of (\ref{tau2}) is
\begin{equation}
\tau_{ii}=2\partial_{X_{i}}\left[  \left(  p-p^{\prime}\right)  \left(
u_{i}-u_{i}^{\prime}\right)  \right]  . \label{taucont}%
\end{equation}

\bigskip\pagebreak

\begin{center}
\textbf{2.1 \ Use of the Navier-Stokes equation\bigskip}
\end{center}

\qquad The Navier-Stokes equation for velocity component $u_{i}(\mathbf{x},t)$
and the incompressibility condition are
\begin{equation}
\partial_{t}u_{i}+\partial_{x_{n}}\left(  u_{i}u_{n}\right)  =-\partial
_{x_{i}}p+\nu\partial_{x_{n}}\partial_{x_{n}}u_{i}\text{ , and\ }%
\partial_{x_{n}}u_{n}=0. \label{NSE}%
\end{equation}
By multiplying the Navier-Stokes equation for $u_{i}$ by $u_{j}^{\prime}$, we
obtain an equation having $u_{j}^{\prime}\partial_{t}u_{i}$ as its
time-derivative term. \ We add and subtract eight such equations to obtain the
equation having as its time-derivative term the expression $u_{j}\partial
_{t}u_{i}-u_{j}^{\prime}\partial_{t}u_{i}+u_{j}^{\prime}\partial_{t}%
u_{i}^{\prime}-u_{i}\partial_{t}u_{j}^{\prime}+u_{i}^{\prime}\partial_{t}%
u_{j}-u_{i}^{\prime}\partial_{t}u_{j}+u_{i}\partial_{t}u_{j}-u_{j}\partial
_{t}u_{i}^{\prime}$ $=\partial_{t}\left[  \left(  u_{i}-u_{i}^{\prime}\right)
\left(  u_{j}-u_{j}^{\prime}\right)  \right]  $.\ \ Algebra is used to
simplify the terms in the resultant equation, and zero is added to the
equation (for convenience) in the form of $\partial_{x_{n}}\left(
u_{i}^{\prime}u_{j}^{\prime}u_{n}\right)  +\partial_{x_{n}^{\prime}}\left(
u_{i}u_{j}u_{n}^{\prime}\right)  $ (which vanishes by incompressibility). \ We
thereby obtain%

\begin{equation}
\partial_{t}d_{ij}+\partial_{x_{n}}\left(  d_{ij}u_{n}\right)  +\partial
_{x_{n}^{\prime}}\left(  d_{ij}u_{n}^{\prime}\right)  =-\tau_{ij}+\nu\left(
\partial_{x_{n}}\partial_{x_{n}}d_{ij}+\partial_{x_{n}^{\prime}}%
\partial_{x_{n}^{\prime}}d_{ij}\right)  . \label{exact 1}%
\end{equation}

\qquad Use of (\ref{derivs}) in (\ref{exact 1}), and use of the identity
$\partial_{x_{n}}\partial_{x_{n}}\left(  fg\right)  =f\partial_{x_{n}}%
\partial_{x_{n}}g+g\partial_{x_{n}}\partial_{x_{n}}f+2\left(  \partial_{x_{n}%
}f\right)  \left(  \partial_{x_{n}}g\right)  $ to simplify the terms
proportional to $\nu$ gives
\begin{equation}
\partial_{t}d_{ij}+\partial_{X_{n}}\digamma_{ijn}+\partial_{r_{n}}%
d_{ijn}=-\tau_{ij}+2\nu\left(  \partial_{r_{n}}\partial_{r_{n}}d_{ij}+\frac
{1}{4}\partial_{X_{n}}\partial_{X_{n}}d_{ij}-e_{ij}\right)  . \label{exact2}%
\end{equation}
As a check, one sees that (\ref{exact2}) is the same as can be obtained by
specializing, for the present case, equation (2.13) in Hill (2001). \ The
trace of (\ref{exact2}) and substitution of (\ref{eii}) and (\ref{taucont})
give
\begin{equation}
\partial_{t}d_{ii}+\partial_{X_{n}}\digamma_{iin}+\partial_{r_{n}}d_{iin}%
=2\nu\partial_{r_{n}}\partial_{r_{n}}d_{ii}-2\left(  \varepsilon
+\varepsilon^{\prime}\right)  +w, \label{exact2trace}%
\end{equation}
\begin{equation}
\text{where \ \ \ \ }w\equiv-2\partial_{X_{n}}\left[  \left(  p-p^{\prime
}\right)  \left(  u_{n}-u_{n}^{\prime}\right)  \right]  +\frac{\nu}{2}%
\partial_{X_{n}}\partial_{X_{n}}d_{ii}-2\nu\partial_{X_{n}}\partial_{X_{n}%
}\left(  p+p^{\prime}\right)  . \label{ww}%
\end{equation}
The limit $r\rightarrow0$ applied to (\ref{exact2trace}) recovers the
definition of $\varepsilon$ in (\ref{strain}). \ It is significant that
$\varepsilon$\ appears in the unaveraged exact equation (\ref{exact2trace})
because $\varepsilon$\ will appear in the average of (\ref{exact2}) only for
the locally isotropic case.

\bigskip

\begin{center}
\textbf{2.2 \ Exact Incompressibility Relationships\bigskip}
\end{center}

\qquad Because $\mathbf{x}$ and $\mathbf{x}^{\prime}$\ are independent
variables,\ $\partial_{x_{i}}u_{j}^{\prime}=0$, and $\partial_{x_{i}^{\prime}%
}u_{j}=0$. Then, incompressibility gives: $\partial_{X_{n}}u_{n}=0,$
$\partial_{X_{n}}u_{n}^{\prime}=0$, $\partial_{r_{n}}u_{n}=0$, $\partial
_{r_{n}}u_{n}^{\prime}=0$, $\partial_{X_{n}}\left(  u_{n}-u_{n}^{\prime
}\right)  =0$, $\partial_{r_{n}}\left(  u_{n}-u_{n}^{\prime}\right)  =0$.
\ The combined use of incompressibility and (\ref{identderivs}) gives
\begin{align}
\partial_{r_{n}}\left[  \left(  u_{j}-u_{j}^{\prime}\right)  \left(
u_{n}-u_{n}^{\prime}\right)  \right]   &  =\partial_{X_{n}}\left[  \left(
u_{j}+u_{j}^{\prime}\right)  \left(  u_{n}-u_{n}^{\prime}\right)  \right]
/2,\label{secdorderincomp}\\
\partial_{r_{j}}\partial_{r_{n}}\left[  \left(  u_{j}-u_{j}^{\prime}\right)
\left(  u_{n}-u_{n}^{\prime}\right)  \right]   &  =\partial_{X_{j}}%
\partial_{X_{n}}\left[  \left(  u_{j}+u_{j}^{\prime}\right)  \left(
u_{n}+u_{n}^{\prime}\right)  \right]  /4. \label{secdorderincomp2}%
\end{align}

\bigskip

\begin{center}
\textbf{3. \ EXACT AVERAGED EQUATIONS\bigskip}
\end{center}

\bigskip

\begin{center}
\textbf{3.1 \ Exact Equations: \ Ensemble Average\bigskip}
\end{center}

\qquad The ensemble is defined as a set of similar flows. \ An example is a
set of mechanically identical wind tunnels operated with the same forcing.
\ Points $\mathbf{x}$ and $\mathbf{x}^{\prime}$ are defined in each flow
relative to the mechanical structures or relative to the corresponding
locations where the flow is (or was) forced. \ Time $t$ is defined for each
flow from the start of the forcing. \ Thus, the space-time points $\left(
\mathbf{x,x}^{\prime},t\right)  $, or equivalently $\left(  \mathbf{X,r}%
,t\right)  $, are in complete correspondence between flows in the ensemble.
\ The ensemble average is defined at each point $\left(  \mathbf{X,r}%
,t\right)  $ as the arithmetical average over the ensemble. \ We denote the
ensemble average by angle brackets $\left\langle \circ\right\rangle _{E}$,
where the subscript $E$\ is a mnemonic for `ensemble.' \ Define the following statistics:%

\begin{align}
D_{ij}\left(  \mathbf{X,r},t\right)   &  \equiv\left\langle d_{ij}%
\right\rangle _{E},\text{ \ }D_{ijn}\left(  \mathbf{X,r},t\right)
\equiv\left\langle d_{ijn}\right\rangle _{E},\text{ \ }T_{ij}\left(
\mathbf{X,r},t\right)  \equiv\left\langle \tau_{ij}\right\rangle
_{E},\nonumber\\
E_{ij}\left(  \mathbf{X,r},t\right)   &  \equiv\left\langle e_{ij}%
\right\rangle _{E},\text{ \ }W\left(  \mathbf{X,r},t\right)  \equiv
\left\langle w\right\rangle _{E},\text{ \ }F_{ijn}\left(  \mathbf{X,r}%
,t\right)  \equiv\left\langle \digamma_{ijn}\right\rangle _{E}. \label{effave}%
\end{align}
The argument list $\left(  \mathbf{X,r},t\right)  $ is shown above to
emphasize that the average applies to the general case of nonstationary,
inhomogeneous turbulence, and that the ensemble average does not eliminate
dependence on any independent variable. \ The argument list $\left(
\mathbf{X,r},t\right)  $ is suppressed where clarity does not suffer.
\ Defining the symbols $D_{ij}$, $D_{ijn}$, $T_{ij}$, $E_{ij}$, $W$, and
$F_{ijn}$\ causes brief notation in later sections. \ Because the ensemble
average is a summation, it commutes with differential operators, and the
average of (\ref{exact2}) is therefore
\begin{equation}
\partial_{t}D_{ij}+\partial_{X_{n}}F_{ijn}+\partial_{r_{n}}D_{ijn}%
=-T_{ij}+2\nu\left[  \partial_{r_{n}}\partial_{r_{n}}D_{ij}+\frac{1}%
{4}\partial_{X_{n}}\partial_{X_{n}}D_{ij}-E_{ij}\right]  . \label{exactave}%
\end{equation}
The average of (\ref{exact2trace}) is
\begin{equation}
\partial_{t}D_{ii}+\partial_{X_{n}}F_{iin}+\partial_{r_{n}}D_{iin}%
=2\nu\partial_{r_{n}}\partial_{r_{n}}D_{ii}-2\left\langle \varepsilon
+\varepsilon^{\prime}\right\rangle _{E}+W, \label{exactrace}%
\end{equation}
\begin{equation}
\text{where \ \ \ \ \ \ \ \ }W\equiv-2\partial_{X_{n}}\left\langle \left(
p-p^{\prime}\right)  \left(  u_{n}-u_{n}^{\prime}\right)  \right\rangle
_{E}+\frac{\nu}{2}\partial_{X_{n}}\partial_{X_{n}}D_{ii}-2\nu\partial_{X_{n}%
}\partial_{X_{n}}\left\langle p+p^{\prime}\right\rangle _{E}. \label{W}%
\end{equation}

\qquad Exact incompressibility conditions on the second-order velocity
structure function are given by the average of (\ref{secdorderincomp}) and
(\ref{secdorderincomp2}) as
\begin{align}
\partial_{r_{n}}D_{jn}  &  =\partial_{X_{n}}\left\langle \left(  u_{j}%
+u_{j}^{\prime}\right)  \left(  u_{n}-u_{n}^{\prime}\right)  \right\rangle
_{E}/2,\label{avesecdinc1}\\
\partial_{r_{j}}\partial_{r_{n}}D_{jn}  &  =\partial_{X_{j}}\partial_{X_{n}%
}\left\langle \left(  u_{j}+u_{j}^{\prime}\right)  \left(  u_{n}+u_{n}%
^{\prime}\right)  \right\rangle _{E}/4. \label{avesecdinc2}%
\end{align}

\bigskip

\begin{center}
\textbf{3.2 \ Exact Equations: \ Temporal Average\bigskip}
\end{center}

\qquad The ensemble average used above is important because it allows us to
simultaneously investigate rapid temporal variation that a temporal average
would smooth and to investigate sharp spatial variation that a spatial average
would smooth. \ It is important to consider temporal and spatial averages
because they are typical of experiments and DNS, respectively. \ Of course, an
ensemble average can be approximated by widely separated temporal or spatial
sampling for stationary or homogeneous turbulence, respectively. \ However,
nearly continuous sampling is typical. \ Thus, we represent the temporal and
spatial averages by integrals, but all results are valid for the sum of
discrete points as well. \ The temporal average is most meaningful when the
turbulence is nearly stationary, and the spatial average is most meaningful
for nearly homogeneous turbulence.

\qquad Let $t_{0}$ be the start time of the temporal average of duration $T$.
\ The operator effecting the temporal average of any quantity $Q$ is denoted
by $\left\langle \circ\right\rangle _{T}$, which has argument list $\left(
\mathbf{X,r,}t_{0},T\right)  $; that is,
\begin{equation}
\left\langle Q\right\rangle _{T}\equiv\frac{1}{T}\int_{t_{0}}^{t_{0}%
+T}Q\left(  \mathbf{X,r},t\right)  dt. \label{timeavedef}%
\end{equation}
For brevity the argument list $\left(  \mathbf{X,r,}t_{0},T\right)  $ is
suppressed where clarity does not suffer. \ The temporal average of
(\ref{exact2}) is
\begin{align}
\left\langle \partial_{t}d_{ij}\right\rangle _{T}+\partial_{X_{n}}\left\langle
\digamma_{ijn}\right\rangle _{T}+\partial_{r_{n}}\left\langle d_{ijn}%
\right\rangle _{T}  &  =-\left\langle \tau_{ij}\right\rangle _{T}\nonumber\\
&  +2\nu\left(  \partial_{r_{n}}\partial_{r_{n}}\left\langle d_{ij}%
\right\rangle _{T}+\frac{1}{4}\partial_{X_{n}}\partial_{X_{n}}\left\langle
d_{ij}\right\rangle _{T}-\left\langle e_{ij}\right\rangle _{T}\right)  .
\label{tempexact2}%
\end{align}
The temporal average of (\ref{exact2trace}) is
\begin{equation}
\left\langle \partial_{t}d_{ii}\right\rangle _{T}+\partial_{X_{n}}\left\langle
\digamma_{iin}\right\rangle _{T}+\partial_{r_{n}}\left\langle d_{iin}%
\right\rangle _{T}=2\nu\partial_{r_{n}}\partial_{r_{n}}\left\langle
d_{ii}\right\rangle _{T}-2\left\langle \varepsilon+\varepsilon^{\prime
}\right\rangle _{T}+\left\langle w\right\rangle _{T}, \label{tempexactrace}%
\end{equation}
\[
\text{where \ \ \ \ }\left\langle w\right\rangle _{T}=-2\partial_{X_{i}%
}\left\langle \left(  p-p^{\prime}\right)  \left(  u_{i}-u_{i}^{\prime
}\right)  \right\rangle _{T}+\frac{\nu}{2}\partial_{X_{n}}\partial_{X_{n}%
}\left\langle d_{ij}\right\rangle _{T}-2\nu\partial_{X_{n}}\partial_{X_{n}%
}\left\langle p+p^{\prime}\right\rangle _{T}.
\]
Now, (\ref{tempexact2}) and (\ref{tempexactrace}) are exact because they are
derived from (\ref{NSE}) without approximations. \ They differ in form from
(\ref{exactave}) and (\ref{exactrace}) only in that the time derivative does
not commute with the temporal average. \ Thus, (\ref{tempexact2}) contains
$\left\langle \partial_{t}d_{ij}\right\rangle _{T}$ whereas (\ref{exact2})
contains $\partial_{t}D_{ij}\equiv\partial_{t}\left\langle d_{ij}\right\rangle
_{E}$.

\qquad Because the data are taken in the rest frame of the anemometers and
$\partial_{t}$ is the time derivative for that reference frame, it follows
that
\begin{equation}
\left\langle \partial_{t}d_{ij}\right\rangle _{T}\equiv\frac{1}{T}\int_{t_{0}%
}^{t_{0}+T}\partial_{t}d_{ij}dt=\left[  d_{ij}\left(  \mathbf{X,r}%
,t_{0}+T\right)  -d_{ij}\left(  \mathbf{X,r},t_{0}\right)  \right]  /T.
\label{tempavederiv}%
\end{equation}
This shows that it is easy to evaluate $\left\langle \partial_{t}%
d_{ij}\right\rangle _{T}$\ using experimental data because only the first (at
$t=t_{0}$)\ and last (at $t=t_{0}+T$)\ data in the time series are used. \ If
$\left[  d_{ij}\left(  \mathbf{X,r},t_{0}+T\right)  -d_{ij}\left(
\mathbf{X,r},t_{0}\right)  \right]  $ is bounded and its ensemble mean varies
less rapidly than $T$, then we can make $\left\langle \partial_{t}%
d_{ij}\right\rangle _{T}$ as small as we like by allowing $T$ to be very large.

\bigskip

\begin{center}
\textbf{3.3 \ Exact Equations: \ Spatial Average\bigskip}
\end{center}

\qquad Let the spatial average be over a region $\mathbb{R}$ in $\mathbf{X}%
$-space. \ The spatial average of any quantity $Q$ is denoted by $\left\langle
Q\right\rangle _{\mathbb{R}}\left(  \mathbf{r},t,\mathbb{R}\right)  $, and is
defined by
\begin{equation}
\left\langle Q\right\rangle _{\mathbb{R}}\equiv\frac{1}{V}\underset
{\mathbb{R}}{\int\int\int}Q\left(  \mathbf{X,r},t\right)  d\mathbf{X,}
\label{Xvolave}%
\end{equation}
where $V$ is the volume of the space region $\mathbb{R}$. \ For brevity, the
argument list $\left(  \mathbf{r},t,\mathbb{R}\right)  $ is suppressed where
clarity does not suffer. \ The spatial average commutes with $\mathbf{r}$\ and
$t$\ differential and integral operations and with ensemble, time, and
$\mathbf{r}$-space\ averages. \ For the divergence in $\mathbf{X}$ of a vector
$q_{n}$, the divergence theorem relates the volume average to the surface
average; that is,
\begin{equation}
\left\langle \partial_{X_{n}}q_{n}\right\rangle _{\mathbb{R}}\equiv\frac{1}%
{V}\int\int\int\partial_{X_{n}}q_{n}d\mathbf{X}=\frac{S}{V}\left(  \frac{1}%
{S}\int\int\check{N}_{n}q_{n}dS\right)  \equiv\frac{S}{V}\oint_{\mathbf{X}%
_{n}}q_{n}, \label{Xsurfave}%
\end{equation}
where $S$ is the surface area bounding the $\mathbf{X}$-space region
$\mathbb{R}$, $dS$ is the differential of surface area, and $\check{N}_{n}$ is
the unit vector oriented outward and normal to the surface. \ For brevity, the
notation $\oint_{\mathbf{X}_{n}}q_{n}$ is used for the $\mathbf{X}$-space
surface average in (\ref{Xsurfave}).

\qquad The spatial average of (\ref{exact2}) is
\begin{align}
\partial_{t}\left\langle d_{ij}\right\rangle _{\mathbb{R}}+\frac{S}{V}%
\oint_{\mathbf{X}_{n}}\digamma_{ijn}+\partial_{r_{n}}\left\langle
d_{ijn}\right\rangle _{\mathbb{R}}  &  =-\left\langle \tau_{ij}\right\rangle
_{\mathbb{R}}\nonumber\\
&  +2\nu\left(  \partial_{r_{n}}\partial_{r_{n}}\left\langle d_{ij}%
\right\rangle _{\mathbb{R}}+\frac{1}{4}\frac{S}{V}\oint_{\mathbf{X}_{n}%
}\partial_{X_{n}}d_{ij}-\left\langle e_{ij}\right\rangle _{\mathbb{R}}\right)
. \label{spatialaveext2}%
\end{align}
The spatial average of (\ref{exact2trace}) is
\begin{equation}
\partial_{t}\left\langle d_{ii}\right\rangle _{\mathbb{R}}+\frac{S}{V}%
\oint_{\mathbf{X}_{n}}\digamma_{iin}+\partial_{r_{n}}\left\langle
d_{iin}\right\rangle _{\mathbb{R}}=2\nu\partial_{r_{n}}\partial_{r_{n}%
}\left\langle d_{ii}\right\rangle _{\mathbb{R}}-2\left\langle \varepsilon
+\varepsilon^{\prime}\right\rangle _{\mathbb{R}}+\left\langle w\right\rangle
_{\mathbb{R}}, \label{spatialavetrace}%
\end{equation}
\[
\text{where\ \ \ }\left\langle w\right\rangle _{\mathbb{R}}\equiv\frac{S}%
{V}\oint_{\mathbf{X}_{n}}\left[  -2\left(  p-p^{\prime}\right)  \left(
u_{n}-u_{n}^{\prime}\right)  +\frac{\nu}{2}\partial_{X_{n}}d_{ij}-2\nu
\partial_{X_{n}}\left(  p+p^{\prime}\right)  \right]  .
\]

\qquad The spatial average of the\ exact incompressibility condition
(\ref{secdorderincomp}) is
\begin{equation}
\partial_{r_{n}}\left\langle d_{jn}\right\rangle _{\mathbb{R}}=\frac{S}%
{2V}\oint_{\mathbf{X}_{n}}\left(  u_{n}-u_{n}^{\prime}\right)  \left(
u_{j}+u_{j}^{\prime}\right)  , \label{spacavescincop}%
\end{equation}
which is, on the right-hand side, a surface flux of a quantity that depends on
large-scale structures in the flow. \ Similarly, (\ref{secdorderincomp2})
gives
\[
\partial_{r_{j}}\partial_{r_{n}}\left\langle d_{jn}\right\rangle _{\mathbb{R}%
}=\frac{S}{4V}\oint_{\mathbf{X}_{n}}\left[  \partial_{X_{j}}\left(
u_{n}+u_{n}^{\prime}\right)  \right]  \left(  u_{j}+u_{j}^{\prime}\right)  .
\]

Of course, (\ref{spatialaveext2}) and (\ref{spacavescincop}) are exact.

\bigskip

\begin{center}
\textbf{3.4 \ Spatial Average: \ DNS with Periodic Boundary Conditions\bigskip
}
\end{center}

\qquad The spatial average is particularly relevant to DNS. \ DNS that is used
to investigate turbulence at small scales often has periodic boundary
conditions. \ For such DNS, consider the spatial average over the entire DNS
domain.\ \ Contributions to $\oint_{\mathbf{X}_{n}}q_{n}$ from opposite sides
of the averaging volume cancel for that case such that $\oint_{\mathbf{X}_{n}%
}q_{n}=0$ and therefore $\left\langle \partial_{X_{n}}q_{n}\right\rangle
_{\mathbb{R}}=0$. \ In (\ref{spatialaveext2}) we then have $\oint
_{\mathbf{X}_{n}}\digamma_{iin}=0$ and $\oint_{\mathbf{X}_{n}}\partial_{X_{n}%
}d_{ij}=0$. \ In (\ref{spatialavetrace}) we have $\oint_{\mathbf{X}_{n}%
}\digamma_{iin}=0$ and $\left\langle w\right\rangle _{\mathbb{R}}=0$. \ In
(\ref{spacavescincop}), the right-hand side vanishes. \ Thus, in the important
DNS case described above, we have the significant simplification that
\begin{equation}
\partial_{t}\left\langle d_{ij}\right\rangle _{\mathbb{R}}+\partial_{r_{n}%
}\left\langle d_{ijn}\right\rangle _{\mathbb{R}}=-\left\langle \tau
_{ij}\right\rangle _{\mathbb{R}}+2\nu\left(  \partial_{r_{n}}\partial_{r_{n}%
}\left\langle d_{ij}\right\rangle _{\mathbb{R}}-\left\langle e_{ij}%
\right\rangle _{\mathbb{R}}\right)  , \label{spaceDNS}%
\end{equation}
\begin{equation}
\partial_{t}\left\langle d_{ii}\right\rangle _{\mathbb{R}}+\partial_{r_{n}%
}\left\langle d_{iin}\right\rangle _{\mathbb{R}}=2\nu\partial_{r_{n}}%
\partial_{r_{n}}\left\langle d_{ii}\right\rangle _{\mathbb{R}}-2\left\langle
\varepsilon+\varepsilon^{\prime}\right\rangle _{\mathbb{R}},
\label{spaceDNStrace}%
\end{equation}
and
\begin{equation}
\partial_{r_{n}}\left\langle d_{jn}\right\rangle _{\mathbb{R}}=0.
\label{spaceincomp2}%
\end{equation}
Proof of $\partial_{r_{j}}\left\langle e_{ij}\right\rangle _{\mathbb{R}}=0$
follows: \ Using (\ref{identderivs}) in (\ref{eij}) we have $\partial_{r_{j}%
}e_{ij}=\partial_{X_{j}}\xi_{ij}$, where $\xi_{ij}\equiv\nu\left[  \left(
\partial_{x_{n}}u_{i}\right)  \left(  \partial_{x_{n}}u_{j}\right)  -\left(
\partial_{x_{n}^{\prime}}u_{i}^{\prime}\right)  \left(  \partial
_{x_{n}^{\prime}}u_{j}^{\prime}\right)  \right]  $ such that $\partial_{r_{j}%
}\left\langle e_{ij}\right\rangle _{\mathbb{R}}=\left\langle \partial_{r_{j}%
}e_{ij}\right\rangle _{\mathbb{R}}=\left\langle \partial_{X_{j}}\xi
_{ij}\right\rangle _{\mathbb{R}}=\frac{S}{V}\oint_{\mathbf{X}_{n}}\xi_{ij}$;
this surface integral\ vanishes because of the DNS periodic boundary
conditions and the selected averaging volume. \ Thus,
\begin{equation}
\partial_{r_{n}}\left\langle e_{jn}\right\rangle _{\mathbb{R}}=0.
\label{spaceincomp3}%
\end{equation}

\qquad No approximations have been used to obtain these equations for the DNS
case considered. \ It seems that (\ref{spaceDNS})-(\ref{spaceDNStrace}) offer
an ideal opportunity to evaluate the contribution of the time-derivative term
$\partial_{t}\left\langle d_{ij}\right\rangle _{\mathbb{R}}$ for freely
decaying turbulence, as well as the contribution of the pressure term
$\left\langle \tau_{ij}\right\rangle _{\mathbb{R}}$ for anisotropic
turbulence, as well as the balance of the off-diagonal components of
(\ref{spaceDNS}).

\qquad Because we have not introduced a force generating the turbulence and
because every point in the flow enters into the $\mathbf{X}$-space average,
the DNS must be freely decaying. \ As shown in Appendix A, it is
straightforward to include forces in our equations.

\qquad Performing the $\mathbf{r}$-space divergence of (\ref{spaceDNS}) and
using (\ref{spaceincomp2})-(\ref{spaceincomp3}), we have
\begin{equation}
\partial_{r_{j}}\partial_{r_{n}}\left\langle d_{ijn}\right\rangle
_{\mathbb{R}}=-\partial_{r_{j}}\left\langle \tau_{ij}\right\rangle
_{\mathbb{R}}. \label{aaa}%
\end{equation}
This exact result is analogous to the asymptotic result in Table 3 of Hill (1997).

\qquad We can further simplify the dissipation-rate term in
(\ref{spaceDNStrace}). \ Using Taylor's series, we have $\varepsilon\left(
\mathbf{x},t\right)  =\varepsilon\left(  \mathbf{X},t\right)  +\frac{r_{n}}%
{2}\partial_{X_{n}}\varepsilon\left(  \mathbf{X},t\right)  +\frac{1}{2}%
\frac{r_{n}}{2}\frac{r_{p}}{2}\partial_{X_{n}}\partial_{X_{p}}\varepsilon
\left(  \mathbf{X},t\right)  +\cdots$. Clearly, a great number of terms will
be needed when $\left|  \mathbf{x-X}\right|  $ is outside of the viscous
range, but the differentiability of hydrodynamic fields guarantees convergence
of the Taylor series. \ The series for $\varepsilon\left(  \mathbf{x}^{\prime
},t\right)  $ is the same as for $\varepsilon\left(  \mathbf{x},t\right)
$\ in which $\mathbf{r}/2$ is replaced by $-\mathbf{r}/2$, such that
\begin{equation}
\varepsilon+\varepsilon^{\prime}=2\varepsilon\left(  \mathbf{X},t\right)
+\frac{1}{4}r_{n}r_{p}\partial_{X_{n}}\partial_{X_{p}}\varepsilon\left(
\mathbf{X},t\right)  +\ldots=2\varepsilon\left(  \mathbf{X},t\right)
+\partial_{X_{n}}\left[  \frac{1}{4}r_{n}r_{p}\partial_{X_{p}}\varepsilon
\left(  \mathbf{X},t\right)  +\ldots\right]  . \label{Taylor seriesdissip}%
\end{equation}
Only terms having even-order derivatives appear in (\ref{Taylor seriesdissip}%
). \ The right-most term in (\ref{Taylor seriesdissip}) has the form
$\partial_{X_{n}}q_{n}$, and therefore vanishes when averaged in $\mathbf{X}%
$-space over the entire DNS domain for the periodic DNS case considered.
\ Substituting (\ref{Taylor seriesdissip}) in (\ref{spaceDNStrace}) gives the
term
\begin{equation}
-2\left\langle \varepsilon+\varepsilon^{\prime}\right\rangle _{\mathbb{R}%
}=-4\left\langle \varepsilon\left(  \mathbf{X},t\right)  \right\rangle
_{\mathbb{R}}=-4\left\langle \varepsilon\right\rangle _{\mathbb{R}}\left(
t\right)  . \label{DNSeps}%
\end{equation}
The same method applied to the right-most term in (\ref{spaceDNS}) gives
\begin{equation}
-\left\langle e_{ij}\right\rangle _{\mathbb{R}}=-4\nu\left\langle \left[
\left(  \partial_{x_{n}}u_{i}\right)  \left(  \partial_{x_{n}}u_{j}\right)
\right]  _{\mathbf{x}=\mathbf{X}}\right\rangle _{\mathbb{R}}\left(  t\right)
\text{,} \label{DNSeij}%
\end{equation}
where the subscript $\mathbf{x}=\mathbf{X}$ means that the derivatives are
evaluated at the point $\mathbf{X}$. \ Of course, none of the quantities in
(\ref{spaceDNS})-(\ref{DNSeij}) depends on $\mathbf{X}$ because of the spatial
average over $\mathbf{X}$. \ An interesting feature of (\ref{DNSeps}) and
(\ref{DNSeij}) is that their right-hand sides clearly do not depend on
$\mathbf{r}$, whereas this is not obvious in (\ref{spaceDNS}) and
(\ref{spaceDNStrace}). \ The only dependence of (\ref{DNSeps}) and
(\ref{DNSeij}) is on $t$. \ Thus, $\left(  t\right)  $ on the right-hand side
of (\ref{DNSeps})-(\ref{DNSeij}) is the entire argument list.

\qquad Of course, these results follow from the periodic boundary conditions
and the fact that the averaging volume is over the whole periodic structure of
the DNS domain. \ These results follow from the symmetry of that case.

\bigskip

\begin{center}
\textbf{4. \ AVERAGES OVER THE }$\mathbf{r}$\textbf{-SPACE SPHERE\bigskip}

\textbf{4.1 \ Definition of the }$\mathbf{r}$\textbf{-Space Sphere
Average\ and the Orientation Average\bigskip}
\end{center}

\qquad The energy dissipation rate averaged over a sphere in $\mathbf{r}%
$-space has been a recurrent theme in small-scale similarity theories since
its introduction by Obukhov (1962) and Kolmogorov (1962). \ By averaging our
equations for the trace, we can, for the first time, produce an exact
dynamical equation containing the sphere-averaged energy dissipation rate.
\ The volume average over an $\mathbf{r}$-space sphere of radius $r_{S}$\ of a
quantity $Q$\ is denoted by
\begin{equation}
\left\langle Q\right\rangle _{\mathbf{r}\text{-sphere}}\equiv\left(
4\pi\left(  r_{S}\right)  ^{3}/3\right)  ^{-1}\underset{\left\vert
\mathbf{r}\right\vert \text{ }\leq\text{ }r_{S}}{\int\int\int}Q\left(
\mathbf{X,r},t\right)  d\mathbf{r.} \label{r-sphere}%
\end{equation}
The orientation average over the surface of the $\mathbf{r}$-space sphere of
radius $r_{S}$ of a vector $q_{n}\left(  \mathbf{X,r},t\right)  $ is denoted
as follows:
\begin{equation}
\oint_{\mathbf{r}_{n}}q_{n}\equiv\left(  4\pi\left(  r_{S}\right)
^{2}\right)  ^{-1}\underset{\mathbf{r}-\text{sphere}}{\int\int}\frac{r_{n}}%
{r}q_{n}\left(  \mathbf{X,r},t\right)  ds, \label{r-surface}%
\end{equation}
where $ds$ is the differential of surface area, and $r_{n}/r$ is the unit
vector oriented outward and normal to the surface of the $\mathbf{r}$-space
sphere. \ Both $\left\langle Q\right\rangle _{\mathbf{r}\text{-sphere}}$\ and
$\oint_{\mathbf{r}_{n}}q_{n}$\ are functions of \ $\mathbf{X}$, $r_{S}$, and
$t$, but the argument list $\left(  \mathbf{X,}r_{S},t\right)  $ is
suppressed. \ In this notation, the divergence theorem is
\begin{equation}
\left\langle \partial_{r_{n}}q_{n}\right\rangle _{\mathbf{r}\text{-sphere}%
}=\left(  3/r_{S}\right)  \oint_{\mathbf{r}_{n}}q_{n}\text{.} \label{r-Gauss}%
\end{equation}
Because $\mathbf{r}$, $\mathbf{X}$, and $t$ are independent variables, the
$\mathbf{r}$-space volume and orientation averages commute with time and
$\mathbf{X}$-space averages and with $\mathbf{X}$- and $t$-differential
operators, and, of course, with the ensemble, temporal , and spatial averages
as well. \ For instance, $\left\langle \partial_{t}\left\langle d_{ii}%
\right\rangle _{\mathbb{R}}\right\rangle _{\mathbf{r}\text{-sphere}}%
$=$\partial_{t}\left\langle \left\langle d_{ii}\right\rangle _{\mathbb{R}%
}\right\rangle _{\mathbf{r}\text{-sphere}}$=$\left\langle \left\langle
\partial_{t}d_{ii}\right\rangle _{\mathbf{r}\text{-sphere}}\right\rangle
_{\mathbb{R}}=\partial_{t}\left\langle \left\langle d_{ii}\right\rangle
_{\mathbf{r}\text{-sphere}}\right\rangle _{\mathbb{R}}$, etc.\textbf{\bigskip}

\begin{center}
\textbf{4.2 \ Example of an Equation Operated upon by the }$\mathbf{r}%
$\textbf{-Space Sphere Average\bigskip}
\end{center}

\qquad The $\mathbf{r}$-sphere average (\ref{r-sphere}) can operate on the
structure-function equations (\ref{exactave}), (\ref{exactrace}),
(\ref{tempexact2}), (\ref{tempexactrace}), (\ref{spatialaveext2}),
(\ref{spatialavetrace}), (\ref{spaceDNS}), (\ref{spaceDNStrace}),
(\ref{approxave}), (\ref{approxspacetrace}), (\ref{approxtrace}),
(\ref{tempapproxave}); indeed, it can operate on the unaveraged equations
(\ref{exact2}) and (\ref{exact2trace}) as well. \ These equations have terms
of the form $\partial_{r_{n}}q_{n}$; examples are: \ $q_{n}=\left\langle
d_{ijn}\right\rangle _{\mathbb{R}}$, $\partial_{r_{n}}\left\langle
d_{ii}\right\rangle _{\mathbb{R}}$, $D_{iin}$, $\left\langle d_{ijn}%
\right\rangle _{T}$, $\partial_{r_{n}}\left\langle d_{ii}\right\rangle _{T}$,
etc. \ By means of (\ref{r-Gauss}), the volume average in $\mathbf{r}$-space
of any term of the form $\partial_{r_{n}}q_{n}$ produces the orientation
average of $q_{n}$ within the subject equation. \ After operating on
(\ref{exactrace}) with the volume average in $\mathbf{r}$-space
(\ref{r-sphere}), the right-most term in that equation contains $\left\langle
\left\langle \varepsilon+\varepsilon^{\prime}\right\rangle _{\mathbf{r}%
\text{-sphere}}\right\rangle _{E}$, which is the same as the sphere-averaged
energy dissipation rate defined in the third equations of both Obukhov (1962)
and Kolmogorov (1962) (after multiplication by 2).

\qquad The result of the $\mathbf{r}$-space sphere average of any of our
equations will be clear from operating on the simplest equation, namely,
(\ref{spaceDNStrace}) for the case of periodic DNS. \ The average of
(\ref{spaceDNStrace}) over a sphere in $\mathbf{r}$-space of radius $r_{S}$
and multiplication by $r_{S}/3$ and use of (\ref{DNSeps}) gives
\begin{equation}
\frac{r_{S}}{3}\partial_{t}\left\langle \left\langle d_{ii}\right\rangle
_{\mathbf{r}\text{-sphere}}\right\rangle _{\mathbb{R}}+\oint_{\mathbf{r}_{n}%
}\left\langle d_{iin}\right\rangle _{\mathbb{R}}=2\nu\oint_{\mathbf{r}_{n}%
}\partial_{r_{n}}\left\langle d_{ii}\right\rangle _{\mathbb{R}}-\frac{4r_{S}%
}{3}\left\langle \left\langle \varepsilon\right\rangle _{\mathbf{r}%
\text{-sphere}}\right\rangle _{\mathbb{R}}.\label{simpleDNS}%
\end{equation}
\ The terms have argument list $\left(  r_{S},t\right)  $, but $\left\langle
\left\langle \varepsilon\right\rangle _{\mathbf{r}\text{-sphere}}\right\rangle
_{\mathbb{R}}$\ depends only on $t$. \ Of course, none of the quantities in
(\ref{simpleDNS}) depends on $\mathbf{X}$ because of the $\mathbf{X}$-space
average. \ Despite its simplicity, (\ref{simpleDNS}) has been obtained without
approximations for the freely decaying DNS case considered; (\ref{simpleDNS}%
)\ applies to inhomogeneous and anisotropic DNS having periodic boundary
conditions.\textbf{\bigskip}\pagebreak

\begin{center}
\textbf{4.3 \ Kolmogorov's Equation Derived from the Sphere-Averaged
Equation\bigskip}
\end{center}

\qquad Most readers are familiar with Kolmogorov's (1941a) famous equation
that is valid for locally isotropic turbulence. \ A useful point of reference
is to derive it from (\ref{simpleDNS}). \ An index $1$ denotes projection in
the direction of $\mathbf{r}$ and indices $2$ and $3$ denote orthogonal
directions perpendicular to $\mathbf{r}$. \ For locally isotropic turbulence
we recall that the only nonzero components of $\left\langle d_{ijn}%
\right\rangle _{\mathbb{R}}$ are $\left\langle d_{111}\right\rangle
_{\mathbb{R}}$, $\left\langle d_{221}\right\rangle _{\mathbb{R}}=\left\langle
d_{331}\right\rangle _{\mathbb{R}}$, and of $\left\langle d_{ij}\right\rangle
_{\mathbb{R}}$ are$\ \left\langle d_{11}\right\rangle _{\mathbb{R}}$,
and\ $\left\langle d_{22}\right\rangle _{\mathbb{R}}=\left\langle
d_{33}\right\rangle _{\mathbb{R}}$. \ These components depend only on $r$ such
that there is no distinction in an $\mathbf{r}$-space sphere average between
$r_{S\text{ }}$ and $r$; thus, we simplify the notation by replacing
$r_{S\text{ }}$ with $r$. \ The isotopic-tensor formula for $\left\langle
d_{ijn}\right\rangle _{\mathbb{R}}$ gives $\left\langle d_{iin}\right\rangle
_{\mathbb{R}}=\left(  r_{n}/r\right)  \left(  \left\langle d_{111}%
\right\rangle _{\mathbb{R}}+2\left\langle d_{221}\right\rangle _{\mathbb{R}%
}\right)  =\left(  r_{n}/r\right)  \left\langle d_{ii1}\right\rangle
_{\mathbb{R}}$, substitution of which into (\ref{r-surface}) gives
$\oint_{r_{n}}\left\langle d_{iin}\right\rangle _{\mathbb{R}}=\left(
r_{n}/r\right)  \left\langle d_{iin}\right\rangle _{\mathbb{R}}=\left(
r_{n}/r\right)  \left(  r_{n}/r\right)  \ \left\langle d_{ii1}\right\rangle
_{\mathbb{R}}=$ \ $\left\langle d_{ii1}\right\rangle _{\mathbb{R}}$. Since
$\left(  \partial_{r_{n}}r\right)  =\left(  r_{n}/r\right)  $, we have
$\oint_{r_{n}}\partial_{r_{n}}\left\langle d_{ii}\right\rangle _{\mathbb{R}%
}=\left(  r_{n}/r\right)  \left(  \partial_{r_{n}}r\right)  \partial
_{r}\left\langle d_{ii}\right\rangle _{\mathbb{R}}=\partial_{r}\left\langle
d_{ii}\right\rangle _{\mathbb{R}}$. \ For locally stationary turbulence, which
is the case considered by Kolmogorov (1941a), the time-derivative term in
(\ref{simpleDNS}) is neglected; then (\ref{simpleDNS}) becomes
\begin{equation}
\left\langle d_{ii1}\right\rangle _{\mathbb{R}}=2\nu\partial_{r}\left\langle
d_{ii}\right\rangle _{\mathbb{R}}-\frac{4}{3}\left\langle \varepsilon
\right\rangle _{\mathbb{R}}r. \label{Kol1trace}%
\end{equation}
Alternatively, we can time average (\ref{simpleDNS}); then the time derivative
can be neglected with the weaker conditions noted with respect to the
smallness of (\ref{tempavederiv}); then
\begin{equation}
\left\langle \left\langle d_{ii1}\right\rangle _{\mathbb{R}}\right\rangle
_{T}=2\nu\partial_{r}\left\langle \left\langle d_{ii}\right\rangle
_{\mathbb{R}}\right\rangle _{T}-\frac{4}{3}\left\langle \left\langle
\varepsilon\right\rangle _{\mathbb{R}}\right\rangle _{T}r. \label{Koltrace}%
\end{equation}
For simplicity of notation, continue with (\ref{Kol1trace}). \ To eliminate
$\left\langle d_{22}\right\rangle _{\mathbb{R}}$ and $\left\langle
\left\langle d_{221}\right\rangle _{\mathbb{R}}\right\rangle $ from the
expressions $\left\langle d_{ii}\right\rangle _{\mathbb{R}}=\left\langle
d_{11}\right\rangle _{\mathbb{R}}+2\left\langle d_{22}\right\rangle
_{\mathbb{R}} $ and $\left\langle d_{ii1}\right\rangle _{\mathbb{R}%
}=\left\langle d_{111}\right\rangle _{\mathbb{R}}+2\left\langle d_{221}%
\right\rangle _{\mathbb{R}}$, we use the incompressibility conditions
$\frac{r}{2}\partial_{r}\left\langle d_{11}\right\rangle _{\mathbb{R}%
}+\left\langle d_{11}\right\rangle _{\mathbb{R}}-\left\langle d_{22}%
\right\rangle _{\mathbb{R}}=0$, and $r\partial_{r}\left\langle d_{111}%
\right\rangle _{\mathbb{R}}+\left\langle d_{111}\right\rangle _{\mathbb{R}%
}-6\left\langle d_{221}\right\rangle _{\mathbb{R}}=0$, which are valid for
local isotropy (Hill, 1997). \ Then (\ref{Kol1trace}) becomes, after
multiplying by $3r^{3}$, $\partial_{r}\left(  r^{4}\left\langle d_{111}%
\right\rangle _{\mathbb{R}}\right)  =6\nu r^{3}\partial_{r}\left[
r^{-2}\partial_{r}\left(  r^{3}\left\langle d_{11}\right\rangle _{\mathbb{R}%
}\right)  \right]  -4\left\langle \varepsilon\right\rangle _{\mathbb{R}}r^{4}%
$, which is then integrated from $0$ to $r$. \ After the term proportional to
$\nu$ is integrated by parts and the resultant equation is divided by $r^{4}$
we have Kolmogorov's equation
\begin{equation}
\left\langle d_{111}\right\rangle _{\mathbb{R}}=6\nu\partial_{r}\left\langle
d_{11}\right\rangle _{\mathbb{R}}-\frac{4}{5}\left\langle \varepsilon
\right\rangle _{\mathbb{R}}r. \label{koleq}%
\end{equation}
Two integrations over $r$\ were required to obtain the equivalent of
(\ref{koleq}) in section 6 of Hill (1997), whereas one integration over
$r$\ was required here to obtain (\ref{koleq}); the reason is that the
$\mathbf{r}$-space sphere average replaced the first integration.
\ Kolmogorov's 4/5 law, $\left\langle d_{111}\right\rangle _{\mathbb{R}%
}=-\frac{4}{5}\left\langle \varepsilon\right\rangle _{\mathbb{R}}r,$ for the
inertial range immediately follows from (\ref{koleq}). \ For the viscous
range, $\left\langle d_{111}\right\rangle _{\mathbb{R}}$ can be neglected in
(\ref{koleq}) such that the known relation $\left\langle \varepsilon
\right\rangle _{\mathbb{R}}=\left(  15\nu/2r\right)  \partial_{r}\left\langle
d_{11}\right\rangle _{\mathbb{R}}=15\nu\left\langle \left(  \partial_{x_{1}%
}u_{1}\right)  ^{2}\right\rangle _{\mathbb{R}}$ is obtained, where the
viscous-range asymptotic formula $\left\langle d_{11}\right\rangle
_{\mathbb{R}}=\left\langle \left(  \partial_{x_{1}}u_{1}\right)
^{2}\right\rangle _{\mathbb{R}}r^{2}$ was used.

\bigskip

\begin{center}
\textbf{5. \ REYNOLDS DECOMPOSITION\bigskip}
\end{center}

\qquad The Reynolds decomposition separates any hydrodynamic variable into its
mean value and fluctuation and is essential when considering hot-wire
anemometer data. \ In the next section, the Reynolds decomposition is used to
elucidate the meaning of $\partial_{X_{n}}F_{ijn}$, and in Sec. 7.4 to perform
the scale analysis.

\qquad For the ensemble average, the Reynolds decomposition of $u_{i}%
(\mathbf{x},t)$ is defined by
\begin{equation}
u_{i}(\mathbf{x},t)\equiv U_{i}(\mathbf{x},t)+\widehat{u}_{i}(\mathbf{x}%
,t)\ ,\text{ where }U_{i}(\mathbf{x},t)\equiv\left\langle u_{i}(\mathbf{x}%
,t)\right\rangle _{E},\text{ and }\left\langle \widehat{u}_{i}(\mathbf{x}%
,t)\right\rangle _{E}=0, \label{defineReydecomp}%
\end{equation}
and similarly at the point $\mathbf{x}^{\prime}$. \ For brevity,
$U_{i}^{\prime}=U_{i}(\mathbf{x}^{\prime},t)$, and $\widehat{u}_{i}^{\prime
}=\widehat{u}_{i}(\mathbf{x}^{\prime},t)$, etc. \ Using (\ref{derivs}), the
incompressibility condition gives
\begin{equation}
\partial_{X_{n}}u_{n}=0,\text{ \ }\partial_{X_{n}}U_{n}=0,\text{ \ }%
\partial_{X_{n}}\widehat{u}_{n}=0,\text{ \ }\partial_{r_{n}}u_{n}=0,\text{
\ }\partial_{r_{n}}U_{n}=0,\text{ \ }\partial_{r_{n}}\widehat{u}_{n}=0,
\label{incompress}%
\end{equation}
and similarly for $u_{n}^{\prime}$, $U_{n}^{\prime}$, $\widehat{u}_{n}%
^{\prime}$.

\qquad For the time average (\ref{timeavedef}), the mean velocity is
$U_{i}(\mathbf{x},t_{0},T)\equiv\left\langle u_{i}(\mathbf{x},t)\right\rangle
_{T}$; as in (\ref{timeavedef}) this notation emphasizes that the mean depends
on the start, $t_{0}$, and duration, $T$, of the time average, as well as on
$\mathbf{x}$. \ The Reynolds decomposition\ is $u_{i}(\mathbf{x},t)\equiv
U_{i}(\mathbf{x},t_{0},T)+\widehat{u}_{i}(\mathbf{x},t,t_{0},T)$, such that
$\left\langle \widehat{u}_{i}(\mathbf{x},t,t_{0},T)\right\rangle _{T}=0$.
\ Clearly the fluctuation, $\widehat{u}_{i}(\mathbf{x},t,t_{0},T)$, also
depends on $t_{0}$ and $T$ as well as on $\mathbf{x}$ and $t$, and
(\ref{incompress}) is valid because the time average commutes with spatial derivatives.

\qquad For the space average (\ref{Xvolave}), it follows from (\ref{change})
that when the integral over $\mathbf{X}$ operates on a single-point quantity
like $u_{i}(\mathbf{x},t)$, it is an integral over $\mathbf{x}$ such that
(\ref{Xvolave}) produces a function only of $t$ but not of $\mathbf{X}$ or
$\mathbf{r}$. \ Thus, the mean velocity is $U_{i}(t,\mathbb{R})\equiv
\left\langle u_{i}(\mathbf{x},t)\right\rangle _{\mathbb{R}}$; as
in\ (\ref{Xvolave}) this average depends on the centroid and shape of the
averaging volume, but this dependence is not denoted explicitly.\ The Reynolds
decomposition\ is $u_{i}(\mathbf{x},t)\equiv U_{i}(t,\mathbb{R})+\widehat
{u}_{i}(\mathbf{x},t,\mathbb{R})$, which gives $\left\langle \widehat{u}%
_{i}(\mathbf{x},t,\mathbb{R})\right\rangle _{\mathbb{R}}=0$. \ Clearly,
(\ref{incompress}) is valid for the space average.

\qquad For brevity, the arguments of mean quantities are not shown in the following.

\bigskip

\begin{center}
\textbf{6. \ MEANING OF THE TERM} $\partial_{X_{n}}F_{ijn}$\textbf{\bigskip}
\end{center}

\qquad The Reynolds decomposition (\ref{defineReydecomp}) used in the second
term of (\ref{exactave}) (i.e., $\partial_{X_{n}}F_{ijn}$) combined with
(\ref{incompress}) gives%

\begin{equation}
\partial_{X_{n}}F_{ijn}=\frac{U_{n}+U_{n}^{\prime}}{2}\partial_{X_{n}}%
D_{ij}+\partial_{X_{n}}\left(  \Delta_{i}\widehat{\Gamma}_{jn}+\Delta
_{j}\widehat{\Gamma}_{in}+\widehat{\Gamma}_{ijn}\right)  \label{reyn}%
\end{equation}
where, for brevity, we define\
\begin{equation}
\Delta_{i}\equiv\left(  U_{i}-U_{i}^{\prime}\right)  ,\text{ }\widehat{\Gamma
}_{in}\equiv\left\langle \left(  \widehat{u}_{i}-\widehat{u}_{i}^{\prime
}\right)  \frac{\widehat{u}_{n}+\widehat{u}_{n}^{\prime}}{2}\right\rangle
_{E},\text{ }\widehat{\Gamma}_{ijn}\equiv\left\langle \left(  \widehat{u}%
_{i}-\widehat{u}_{i}^{\prime}\right)  \left(  \widehat{u}_{j}-\widehat{u}%
_{j}^{\prime}\right)  \frac{\widehat{u}_{n}+\widehat{u}_{n}^{\prime}}%
{2}\right\rangle _{E}. \label{gammadef}%
\end{equation}
Note that $\widehat{\circ}$ means that the quantity is a fluctuation, e.g.
$\widehat{u}_{i}$, and that a statistic is calculated from fluctuations, e.g.,
$\widehat{\Gamma}_{in}$\ and $\widehat{D}_{ij}$. \ Also, $D_{ij}$\ appears in
(\ref{reyn}), not $\widehat{D}_{ij}$.

\qquad Consider the first term in (\ref{reyn}), namely $\frac{1}{2}\left(
U_{n}+U_{n}^{\prime}\right)  \partial_{X_{n}}D_{ij}$. \ If the mean flow is
spatially uniform to the extent that $U_{n}$\ and $U_{n}^{\prime}$\ are equal,
then $\frac{1}{2}\left(  U_{n}+U_{n}^{\prime}\right)  \partial_{X_{n}}D_{ij}$
becomes the same expression that Lindborg (1999) [his Eq.(8)] deduced as an
addition to Kolmogorov's equation. \ His deduction was based on Galilean
invariance applied to a uniform mean flow. \ The combination of (\ref{reyn})
and (\ref{exactave}) shows that both $\partial_{t}D_{ij}$\ and $\frac{1}%
{2}\left(  U_{n}+U_{n}^{\prime}\right)  \partial_{X_{n}}D_{ij}$\ must appear
in the dynamical equation as was correctly deduced by Lindborg (1999) on the
basis of mean-flow Galilean invariance, but replacing $\partial_{t}D_{ij}$
with $\frac{1}{2}\left(  U_{n}+U_{n}^{\prime}\right)  \partial_{X_{n}}D_{ij}$,
as was done by Danaila \textit{et al.} (1999 a,b) on the basis of Taylor's
hypothesis, does not preserve that invariance. \ Now, $\partial_{X_{n}}D_{ij}$
is a measure of inhomogeneity because $\partial_{X_{n}}D_{ij}$ is the rate of
change of $D_{ij}\left(  \mathbf{X,r},t\right)  $ with respect to where the
average is performed. \ Thus, $\frac{1}{2}\left(  U_{n}+U_{n}^{\prime}\right)
\partial_{X_{n}}D_{ij}$ describes the effect of the fluid moving relative to
the anemometers\ in a direction in which $D_{ij}\left(  \mathbf{X,r},t\right)
$\ is inhomogeneous. \ Lindborg (1999) quantifies the contribution of
$\frac{1}{2}\left(  U_{n}+U_{n}^{\prime}\right)  \partial_{X_{n}}D_{ij}$ to
Kolmogorov's (1941a) equation (Sec. 4.3) for several experiments and thereby
shows that the contribution can be significant.

\qquad Now, $\frac{1}{2}\left(  U_{n}+U_{n}^{\prime}\right)  \partial_{X_{n}%
}D_{ij}$ is well illustrated by the case of turbulent flow in a pipe or wind
tunnel. \ Perform the $\mathbf{X}$-space spatial average (\ref{Xvolave})\ of
$\frac{1}{2}\left(  U_{n}+U_{n}^{\prime}\right)  \partial_{X_{n}}D_{ij} $ over
a cylinder having sides parallel to the mean velocity and having ends
perpendicular to the mean velocity. \ For simplicity, assume that the mean
velocity is uniform over the ends of the cylinder so that $U_{n}^{\prime
}=U_{n}=\left|  \mathbf{U}\right|  \check{s}_{n}$ where $\check{s}_{n}$ is a
unit vector in the streamwise direction, which is\ the 1-axis.\ \ Use of the
divergence theorem (\ref{Xsurfave}) gives
\begin{align}
\frac{1}{V}\int\int\int\partial_{X_{n}}\left[  \frac{1}{2}\left(  U_{n}%
+U_{n}^{\prime}\right)  D_{ij}\right]  d\mathbf{X}  &  =\frac{1}{A\text{\L }%
}\int\int\check{N}_{n}\check{s}_{n}\left|  \mathbf{U}\right|  D_{ij}%
dS\nonumber\\
&  =\frac{\left|  \mathbf{U}\right|  }{\text{\L }}\left[  \oint_{\mathbf{X}%
_{n}}\check{s}_{n}D_{ij\text{downstream}}-\oint_{\mathbf{X}_{n}}\check{s}%
_{n}D_{ij\text{upstream}}\right]  , \label{updown}%
\end{align}
where $\oint_{\mathbf{X}_{n}}\check{s}_{n}D_{ij\text{downstream}}$ and
$\oint_{\mathbf{X}_{n}}\check{s}_{n}D_{ij\text{upstream}}$ are the surface
averages over just the downstream and upstream ends of the cylinder,
respectively, and \L , $A$,\ and $V=A$\L \ are the length, area of the ends,
and volume of the cylinder, respectively.\ \ Now, $\left(  \left|
\mathbf{U}\right|  /\text{\L }\right)  ^{-1}$\ is the mean time for the flow
to pass from the upstream end of the cylinder to the downstream end. \ Thus,
(\ref{updown}) is the rate of downstream decay of $D_{ij}$ averaged over the
cylinder cross section.

\qquad Now consider the term $\partial_{X_{n}}\left(  \Delta_{i}%
\widehat{\Gamma}_{jn}+\Delta_{j}\widehat{\Gamma}_{in}+\widehat{\Gamma}%
_{ijn}\right)  $\ in (\ref{reyn}). \ From (\ref{gammadef}) this term is
important if there is strong correlation between velocity difference and
velocity sum. \ One such case is when at least one anemometer is at the edge
of a jet and is therefore sometimes immersed in quiescent entrained fluid and
sometimes in turbulently agitated fluid. \ More generally, the second term in
(\ref{reyn}) is important for the case of large-scale structures. \ This term
describes a contribution caused by inhomogeneity in the direction transverse
to the mean flow direction as well as in the streamwise direction. \ Thus,
this term is expected to contribute for pipe and jet flows when anemometers
are separated transverse to the flow. \ Experimental and/or numerical
evaluation of these terms is needed to quantify their contribution to
(\ref{exactave}) for particular flows.

\qquad On the other hand, the second term\ in (\ref{reyn}), i.e.,
$\partial_{X_{n}}\left(  \Delta_{i}\widehat{\Gamma}_{jn}+\Delta_{j}%
\widehat{\Gamma}_{in}+\widehat{\Gamma}_{ijn}\right)  $,does not grow if
$\frac{1}{2}\left(  U_{n}+U_{n}^{\prime}\right)  $ increases, as does the
first term, i.e., $\frac{U_{n}+U_{n}^{\prime}}{2}\partial_{X_{n}}D_{ij}$.
\ Therefore, for a flow in which large-scale structures are minimized, such as
grid-generated turbulence, and for a large enough Reynolds number such that
$r$ can be much less than the integral scale, the second term in (\ref{reyn})
is expected to be\ negligible because it is $\partial_{X_{n}}$\ operating on
an average. \ For such a flow, one expects that the two-point sum, $\left(
\widehat{u}_{n}+\widehat{u}_{n}^{\prime}\right)  $, has a weak statistical
relationship to the difference, $\left(  \widehat{u}_{i}-\widehat{u}%
_{i}^{\prime}\right)  $. \ The negligibility of the second term in
(\ref{reyn}) when (\ref{reyn}) is substituted in (\ref{exactave}) will be
considered further in Sec. 7.4.

\bigskip

\begin{center}
\textbf{7. \ APPROXIMATE EQUATIONS PERTAINING TO EXPERIMENTS ON THE
SMALL-SCALE STRUCTURE OF HIGH-REYNOLDS-NUMBER TURBULENCE\bigskip}
\end{center}

\qquad We are now in a position to investigate three closely related
objectives that will be considered simultaneously. \ One\ objective is to
study the simplification of (\ref{exactave}) on the basis of data for the
small-scale structure of high-Reynolds-number turbulence; another is to
determine the approximations required for that simplification. \ The third
objective is to obtain from (\ref{exactave}) an equation that is closer to the
measurement process of extracting a mean velocity from anemometry data. \ We
use the ensemble-averaged equations because they retain both temporal and
spatial variability. \ Here, we consider the approach toward local
homogeneity. \ For this purpose, our equations that depend on the location of
measurements, i.e., $\mathbf{X}$, are needed. \ We also consider the approach
toward local stationarity, so the dependence on $t$ is needed. \ The
restrictions required by local isotropy are not used, so dependence on the
orientation of measurement, i.e., $\mathbf{r}/r$, is retained. \ On the other
hand, assumptions about the order of magnitude of some quantities require that
local isotropy is not greatly violated. \ The data used for this investigation
are given in Appendix B, which includes the empirically verified (Monin and
Yaglom, 1975) formulas for the inertial and viscous ranges for components of
$D_{ij}$ and $D_{ijn}$.

\bigskip

\begin{center}
\textbf{7.1 \ Structure Functions of Fluctuations\bigskip}
\end{center}

\qquad An experimenter usually extracts $U_{i}$ from the anemometer's signal,
then calculates statistics from $\widehat{u}_{i}$, e.g., $\widehat{D}%
_{ij}\equiv\left\langle \left(  \widehat{u}_{i}-\widehat{u}_{i}^{\prime
}\right)  \left(  \widehat{u}_{j}-\widehat{u}_{j}^{\prime}\right)
\right\rangle _{E}.$ \ Similarly define $\widehat{D}_{ijn}$, $\widehat{T}%
_{ij}$, and $\widehat{E}_{ij}$ in terms of the fluctuations of velocity and
pressure. \ However, $\left(  u_{n}+u_{n}^{\prime}\right)  /2$ in
(\ref{effunav}) cannot be replaced by $\left(  \widehat{u}_{n}+\widehat{u}%
_{n}^{\prime}\right)  /2$ without destroying the meaning of $F_{ijn}$; that
replacement would result in $\widehat{F}_{ijn}$\ being defined as
$\widehat{\Gamma}_{ijn}$\ in (\ref{gammadef}). \ A reasonable choice for the
symbol $\widehat{F}_{ijn}$\ is
\[
\widehat{F}_{ijn}\equiv\left\langle \frac{U_{n}+U_{n}^{\prime}}{2}\left(
\widehat{u}_{i}-\widehat{u}_{i}^{\prime}\right)  \left(  \widehat{u}%
_{j}-\widehat{u}_{j}^{\prime}\right)  \right\rangle _{E}=\frac{U_{n}%
+U_{n}^{\prime}}{2}\widehat{D}_{ij}.
\]

\qquad Now (\ref{exactave}) is not exactly satisfied by substitution of
$\widehat{D}_{ij}$, $\widehat{F}_{ijn}$, $\widehat{D}_{ijn}$, $\widehat
{T}_{ij}$, and $\widehat{E}_{ij}$,\ in place of $D_{ij}$, $F_{ijn}$, $D_{ijn}%
$, $T_{ij}$, and $E_{ij}$, nor does that substitution satisfy any equations
derived from (\ref{exactave}), Kolmogorov's equation being one such equation
(see Hill, 1997). \ Substitution of the Reynolds decomposition of $D_{ij}$,
$F_{ijn}$, $D_{ijn}$, $T_{ij}$, and $E_{ij}$ (e.g., $D_{ij}=\Delta_{i}%
\Delta_{j}+\widehat{D}_{ij}$, etc.) in (\ref{exactave}) gives a complicated
equation. \ Below, simpler approximate equations are derived by scale analysis
and are summarized in Sec. 8.

\bigskip

\begin{center}
\textbf{7.2 \ Experimentally Evaluatable Exact Incompressibility
Conditions\bigskip}
\end{center}

\qquad Because the approximations $\partial_{r_{n}}D_{in}\simeq0$ and
$\partial_{r_{n}}\widehat{D}_{in}\simeq0$ have an essential role in many
theories, experimental evaluation of these approximations is desirable.
\ However, the expressions $\partial_{r_{n}}D_{in}$ and $\partial_{r_{n}%
}\widehat{D}_{in}$ are nearly impossible to evaluate experimentally. \ Use of
(\ref{identderivs}) and (\ref{incompress})\ gives exact expressions for them
that can be more readily evaluated; namely,
\begin{equation}
\partial_{r_{n}}\widehat{D}_{in}=\left\langle \left[  \partial_{r_{n}}\left(
\widehat{u}_{i}-\widehat{u}_{i}^{\prime}\right)  \right]  \left(  \widehat
{u}_{n}-\widehat{u}_{n}^{\prime}\right)  \right\rangle _{E}=\partial_{X_{n}%
}\left\langle \left(  \widehat{u}_{i}+\widehat{u}_{i}^{\prime}\right)  \left(
\widehat{u}_{n}-\widehat{u}_{n}^{\prime}\right)  \right\rangle _{E}/2,
\label{Dincomp}%
\end{equation}
which is similar to (\ref{avesecdinc1}). \ For the temporal average, the
right-most expression in (\ref{Dincomp}) requires, at most, measurements at
four positions of the statistic $\left\langle \left(  \widehat{u}_{i}%
+\widehat{u}_{i}^{\prime}\right)  \left(  \widehat{u}_{n}-\widehat{u}%
_{n}^{\prime}\right)  \right\rangle _{T}$. \ If, as in the case of
grid-generated turbulence, inhomogeneity is streamwise, then only two
positions displaced in the streamwise direction suffice to determine
$\partial_{X_{1}}\left\langle \left(  \widehat{u}_{i}+\widehat{u}_{i}^{\prime
}\right)  \left(  \widehat{u}_{1}-\widehat{u}_{1}^{\prime}\right)
\right\rangle _{E}/2$. \ The Reynolds decomposition gives $\partial_{r_{n}%
}D_{in}=\Delta_{n}\partial_{r_{n}}\Delta_{i}+\partial_{r_{n}}\widehat{D}_{in}%
$, which shows that evaluation of $\ \partial_{r_{n}}D_{in}$ only requires
mean velocity measurements at several positions in addition to the previous
evaluation of \ $\partial_{r_{n}}\widehat{D}_{in}$.

\bigskip

\begin{center}
\textbf{7.3 \ A Necessary Condition for Local Homogeneity\bigskip}
\end{center}

\qquad We must define several scaling parameters determined by the flow. \ The
integral scale, as traditionally defined, is strictly applicable only to
homogeneous turbulence; see, for example, Tennekes and Lumley (1972). \ Here,
however, we are studying inhomogeneous turbulence. \ As an example of the
difficulty of defining integral scales in general inhomogeneous turbulence,
consider the horizontally homogeneous atmospheric surface during daytime
convective conditions. \ It is difficult to imagine a useful integral scale
defined using data obtained along a line from the ground to the upper reaches
of the surface layer. \ However, the horizontal homogeneity\ and Taylor's
hypothesis allow integral scales to be defined for all three velocity
components measured at a point. \ Using surface-layer data, Kaimal et al.
(1976) show that the horizontal velocity components scale with the depth of
the entire boundary layer; that depth can be 1 to 2 km. \ Unlike the
horizontal velocity component, the vertical velocity variance obeys
Monin-Obukhov similarity such that its integral scale is proportional to the
height above ground (Kaimal et al., 1976). \ For our study of the approach
toward local homogeneity, it is necessary to define the large scale as the
smallest of the integral scales or of the distance to boundaries. \ From the
example of the atmospheric surface layer, that scale is the height above
ground. \ Denote this chosen length scale by $L$\ and call it the outer scale.
\ This name distinquishes it from the integral scale, which might not exist as
traditionally defined in terms of the integral of a velocity correlation
function. \ It is useful to define a velocity scale $\upsilon$ by
\begin{equation}
\upsilon\equiv\left(  \left\langle \varepsilon\right\rangle _{E}L\right)
^{1/3}.\label{velscale}%
\end{equation}
Monin and Yaglom (1975) and Tennekes and Lumley (1972) determine that
$\upsilon$ is an estimate of the root-mean-square velocity, and that the mean
shear is not greater than $\upsilon/L$. \ If this is not so for our chosen
outer scale $L$, then $L$\ can be adjusted to make it so. \ From studies of
nearly homogeneous turbulence, the right-hand side of (\ref{velscale}) is
proportional to velocity variance and the proportionality constant is
independent of Reynolds number at high enough Reynolds numbers (Sreenivasan,
1998; Pearson, Krogstad, and van de Water, 2002). \ The proportionality
constant is of order unity and depends somewhat on the large-scale structure
of the flow (Sreenivasan, 1998; Pearson, Krogstad, and van de Water, 2002).

\qquad We define the scale $\ell$ by
\begin{equation}
\ell\equiv10\eta;\label{elscale}%
\end{equation}
$\ell$ is a scale typical of the energy dissipation range (Appendix B.1).
\ Here, Kolmogorov's microscale $\eta$, which is a scale typical of the
viscous range, is defined by
\[
\eta\equiv\left(  \nu^{3}/\left\langle \varepsilon\right\rangle _{E}\right)
^{1/4}.
\]
If the data have an inertial range, then $\ell$ is closely related to the $r$
at which asymptotic formulas for the inertial and viscous ranges are equal;
this is demonstrated in Appendix B.1.

\qquad The basic tenet of local homogeneity is that as $r$ is reduced relative
to $L$, nonlinear randomization causes statistics of differences of basic
hydrodynamic quantities to decrease their dependence on the large-scale flow
structure. \ For $r<\ell$ and as $r$ is further reduced, the nonlinear
randomization is increasingly opposed by the smoothing effect of viscosity.
\ Therefore, $\ell\ll L$ is a necessary condition for local homogeneity. \ For
$r\geq\ell$, $r\ll L$ is the necessary condition. \ That is, local homogeneity
applies to the asymptotic case:
\begin{equation}
\text{if }r<\ell\text{,\ then }\ell\ll L\text{; if }r>\ell\text{,\ then }r\ll
L\text{; i.e., }\max\left(  r,\ell\right)  \ll L,\label{loho1}%
\end{equation}
where $L$ is the outer scale. \ We study the approach toward local homogeneity
by using (\ref{loho1}) in scale analyses. \ We do so in Sec. 7.4, and find
that some predictions of local homogeneity (such as $\partial_{X_{n}}%
F_{ijn}=0$ and $\partial_{r_{n}}D_{jn}=0$) do not follow solely on the basis
of the necessary condition (\ref{loho1}). \ Thus, (\ref{loho1}) is not a
sufficient condition for local homogeneity.

\qquad Suppose for the moment that the turbulence under investigation is
sufficiently homogenous that an integral scale $L$\ can be defined in terms of
an integral of the velocity correlation function. \ The microscale Reynolds
number (Tennekes and Lumley, 1972) $R_{\lambda}$ is well known to be related
to integral scale $L$ and $\eta$ by $L/\eta\varpropto R_{\lambda}^{3/2}$
(Tennekes and Lumley, 1972). \ Then, (\ref{elscale}) gives $L/\ell\varpropto
R_{\lambda}^{3/2}$.\ \ Now, $\ell\ll L$ is a necessary condition in
(\ref{loho1}); so $R_{\lambda}\gg1$ is a necessary condition for local
homogeneity, but it is not a sufficient condition. \ In a general
inhomogeneous turbulence case, we assume that this is also true when $L$\ is
the outer scale.

\bigskip

\begin{center}
\textbf{7.4 \ Scale Analysis\bigskip}
\end{center}

\qquad This section uses the data given in equations (\ref{inertial2}) to
(\ref{gammapprox}) of Appendix B.2. \ Those equations are distinguished by the
prefix B.

\qquad Now, we consider the scale analysis of (\ref{exactave}). \ First,
consider the Reynolds decomposition of $D_{ij}$. \ Denote the local shear at
point $\mathbf{X}$\ by
\[
G_{i,n}\equiv\partial_{X_{n}}U_{i}(\mathbf{X},t).
\]
On the basis (\ref{loho1}) that $\max\left(  r,\ell\right)  \ll L$, we retain
only the first two terms of the Taylor series of $U_{i}$ and $U_{i}^{\prime}$
around point $\mathbf{X}$ to obtain that $\Delta_{i}\simeq r_{p}%
G_{i,p}=rG_{i,1}$ where the 1-axis is parallel to $\mathbf{r}$. \ Therefore,
$\partial_{r_{n}}\Delta_{i}\simeq\left(  \partial_{r_{n}}r_{p}\right)
G_{i,p}=\delta_{pn}G_{i,p}=G_{i,n}$, which also follows from
(\ref{identderivs}). \ Recall that the velocity scale $\upsilon$ is defined
such that $\left\langle \varepsilon\right\rangle _{E}$ is of order
$\upsilon^{3}/L$ and a component of mean shear, i.e., $G_{i,p}$, is at most of
order $\upsilon/L$. \ The Reynolds decomposition of $D_{ij}$ gives
\begin{equation}
D_{ij}=\Delta_{i}\Delta_{j}+\widehat{D}_{ij}\simeq r^{2}G_{i,1}G_{j,1}%
+\widehat{D}_{ij}. \label{ReynD2}%
\end{equation}
Now $G_{j,1}$ might be zero; if not, it is no greater than of order
$\upsilon/L$. \ Use of (\ref{inertial2}) gives $r^{2}G_{\alpha,1}G_{\alpha
,1}/D_{\alpha\alpha}\sim\left(  r/L\right)  ^{4/3}$ in the inertial range, and
use of (\ref{viscous}) gives $r^{2}G_{\alpha,1}G_{\alpha,1}/D_{\alpha\alpha
}\sim\left(  \ell/L\right)  ^{4/3}$ in the viscous range. \ Thus, on the basis
of (\ref{loho1}), (\ref{ReynD2}) gives $D_{\alpha\alpha}\simeq\widehat
{D}_{\alpha\alpha}.$ \ Therefore, (\ref{inertial2}) and (\ref{viscous}) are
used below for $\widehat{D}_{\alpha\alpha}$ as well as for $D_{\alpha\alpha}$.

Consider the Reynolds decomposition of the term $\partial_{r_{n}}D_{ijn}$ in
(\ref{exactave}). \ Use of incompressibility (\ref{incompress}) gives
\begin{align}
\partial_{r_{n}}D_{ijn}  &  =\Delta_{n}\left(  \partial_{r_{n}}\Delta
_{i}\Delta_{j}\right)  +\left(  \partial_{r_{n}}\Delta_{i}\right)  \widehat
{D}_{nj}+\left(  \partial_{r_{n}}\Delta_{j}\right)  \widehat{D}_{in}%
+\nonumber\\
&  \Delta_{n}\partial_{r_{n}}\widehat{D}_{ij}+\Delta_{i}\partial_{r_{n}%
}\widehat{D}_{nj}+\Delta_{j}\partial_{r_{n}}\widehat{D}_{in}+\partial_{r_{n}%
}\widehat{D}_{ijn}. \label{d3decomp}%
\end{align}
The diagonal components of (\ref{d3decomp}) can be compared with the diagonal
elements of $E_{ij}$, which, according to our data (\ref{ediagonal}), are of
order $\left\langle \varepsilon\right\rangle _{E}$. \ The first term in
(\ref{d3decomp}) can be approximated by $r^{2}G_{n,1}\left(  G_{\alpha
,1}G_{\alpha,n}+G_{\alpha,1}G_{\alpha,n}\right)  $, which is at most of order
$r^{2}\upsilon^{3}/L^{3}$; this is of order $\left(  r/L\right)  ^{2}$
relative to $\left\langle \varepsilon\right\rangle _{E}$. \ Hence, when
(\ref{d3decomp}) is substituted in (\ref{exactave}), the first term in
(\ref{d3decomp}) can be neglected relative to the diagonal element
$E_{\alpha\alpha}$ on the basis of (\ref{loho1}). \ The second, third, and
fourth terms in (\ref{d3decomp}) introduce off-diagonal elements of
$\widehat{D}_{ij}$ into the diagonal elements of (\ref{exactave}). \ Using
(\ref{inertial2}) and the assumption (see Appendix B) that the off-diagonal
elements of $D_{ij}$ are no greater than the $D_{\alpha\alpha}$, the second,
third, and fourth terms in (\ref{d3decomp}) can be shown to be no greater than
of order $\left(  r/L\right)  ^{2/3}$ relative to $\left\langle \varepsilon
\right\rangle _{E},$ and are therefore also neglected on the basis of
(\ref{loho1}). \ The same procedure can be used for the fifth and sixth terms
in (\ref{d3decomp}). \ On the other hand, substitution of the definition
(\ref{gammadef}) of $\widehat{\Gamma}_{nj}$\ in (\ref{Dincomp}) and use of
(\ref{gammapprox})\ gives $\partial_{r_{n}}\widehat{D}_{nj}=\partial_{X_{n}%
}\widehat{\Gamma}_{nj}\leq\upsilon^{2}/L$ such that the fifth and sixth terms
in (\ref{d3decomp}) are much less than $\left(  r/L\right)  \left(
\upsilon^{3}/L\right)  $ and are therefore negligible compared with
$\left\langle \varepsilon\right\rangle _{E}$ on the basis of both
(\ref{gammapprox})\ and (\ref{loho1}). \ Therefore, for the projection of
(\ref{exactave}) in an arbitrary direction $\mathbf{\breve{a}}$,
(\ref{loho1}), incompressibility, and our data imply that $\breve{a}_{i}%
\breve{a}_{j}\partial_{r_{n}}D_{ijn}$ can be replaced by $\breve{a}_{i}%
\breve{a}_{j}\partial_{r_{n}}\widehat{D}_{ijn}$.

\qquad The stronger conclusion that $\breve{a}_{i}\breve{a}_{j}\partial
_{r_{n}}D_{ijn}\simeq\breve{a}_{i}\breve{a}_{j}\partial_{r_{n}}\widehat
{D}_{ijn}$\ can be obtained as follows. \ For an inertial range, the above
comparison of terms with $E_{\alpha\alpha}$ is equivalent to comparison with
$\breve{a}_{i}\breve{a}_{j}\partial_{r_{n}}D_{ijn}$ because of
(\ref{arbproD3div}) and (\ref{ediagonal}). \ Therefore, the above scale
analysis is sufficient to state that for the inertial range $\breve{a}%
_{i}\breve{a}_{j}\partial_{r_{n}}D_{ijn}\simeq$ $\breve{a}_{i}\breve{a}%
_{j}\partial_{r_{n}}\widehat{D}_{ijn}$. We need only extend this result to the
viscous range as follows. \ Use of (\ref{arbproD3divvisc}) shows that the
first term in (\ref{d3decomp}) is of order $\left(  \ell/L\right)  ^{2}$
relative to $\breve{a}_{i}\breve{a}_{j}\partial_{r_{n}}D_{ijn}$ in the viscous
range, and that the second through sixth terms in (\ref{d3decomp}) are of
order $\left(  \ell/L\right)  ^{2/3}$ relative to $\breve{a}_{i}\breve{a}%
_{j}\partial_{r_{n}}D_{ijn}$. \ Therefore, (\ref{loho1}), incompressibility,
and the empirical formulas (\ref{arbproD3div})-(\ref{ediagonal}) give
\begin{equation}
\breve{a}_{i}\breve{a}_{j}\partial_{r_{n}}D_{ijn}\simeq\breve{a}_{i}\breve
{a}_{j}\partial_{r_{n}}\widehat{D}_{ijn}. \label{d3aprox}%
\end{equation}
The significance of there being a projection in an arbitrary direction
$\mathbf{\breve{a}}$ within (\ref{d3aprox}), is that empirical evidence is
lacking for the off-diagonal components of $D_{ijn}$.

\qquad We are now ready to consider in more detail the second term in
(\ref{reyn}), namely $\partial_{X_{n}}\left(  \Delta_{i}\widehat{\Gamma}%
_{jn}+\Delta_{j}\widehat{\Gamma}_{in}+\widehat{\Gamma}_{ijn}\right)  $. \ It
is assumed that our data are chosen to mitigate large-scale structures such
that (\ref{gammapprox}) is true. \ One part of the second term in (\ref{reyn})
is $\partial_{X_{n}}\left[  \Delta_{i}\widehat{\Gamma}_{jn}\right]  =\left(
\partial_{X_{n}}\Delta_{i}\right)  \widehat{\Gamma}_{jn}+\Delta_{i}\left(
\partial_{X_{n}}\widehat{\Gamma}_{jn}\right)  $. \ Now, $\left(
\partial_{X_{n}}\Delta_{i}\right)  \simeq r\partial_{X_{n}}G_{i,1}$; this is
at most of order $\left(  r/L\right)  \left(  \upsilon/L\right)  $.
$\ $Therefore, the ratio $\left[  \left(  \partial_{X_{n}}\Delta_{\alpha
}\right)  \widehat{\Gamma}_{\alpha n}\right]  /E_{\alpha\alpha}$ is at most of
order $\left(  r/L\right)  \left(  \widehat{\Gamma}_{\alpha n}/\upsilon
^{2}\right)  $, which is very small compared to unity on the basis of
(\ref{loho1}) and (\ref{gammapprox}). \ Similarly, $\Delta_{\alpha}\left(
\partial_{X_{n}}\widehat{\Gamma}_{\alpha n}\right)  /E_{\alpha\alpha}$ is of
order $\left(  r/L\right)  \left(  \widehat{\Gamma}_{\alpha n}/\upsilon
^{2}\right)  $. \ Another part of the second term in (\ref{reyn}) is
$\partial_{X_{n}}\widehat{\Gamma}_{ijn}$. \ The ratio $\left(  \partial
_{X_{n}}\widehat{\Gamma}_{\alpha\alpha n}\right)  /E_{\alpha\alpha}$ is at
most of order $\widehat{\Gamma}_{\alpha\alpha n}/\upsilon^{3}$, which is very
small because of (\ref{gammapprox}). \ Therefore, the entire second term in
(\ref{reyn}) is negligible compared to $E_{\alpha\alpha}$, and therefore it is
negligible in diagonal components of (\ref{exactave}). \ Neglecting the second
term in (\ref{reyn}) and using incompressibility, in the diagonal components
of (\ref{exactave}) we have
\begin{equation}
\partial_{X_{n}}F_{\alpha\alpha n}\left(  \mathbf{X,r},t\right)  \simeq
\frac{1}{2}\left(  U_{n}+U_{n}^{\prime}\right)  \partial_{X_{n}}%
D_{\alpha\alpha}. \label{effhighR}%
\end{equation}
The Reynolds decomposition of (\ref{effhighR}) is
\begin{equation}
\partial_{X_{n}}F_{\alpha\alpha n}\simeq M_{\alpha\alpha}+\frac{1}{2}\left(
U_{n}+U_{n}^{\prime}\right)  \partial_{X_{n}}\widehat{D}_{\alpha\alpha},
\label{effreyn}%
\end{equation}
\[
\text{where }M_{\alpha\alpha}\equiv\frac{1}{2}\left(  U_{n}+U_{n}^{\prime
}\right)  r^{2}\partial_{X_{n}}\left(  G_{\alpha,1}G_{\alpha,1}\right)  .
\]
There are clearly flows for which we expect that $M_{\alpha\alpha}$\ is
negligible; an example is freely decaying grid-generated turbulence in a wind
tunnel for which $G_{\alpha,1}=0$. \ On the other hand, $M_{\alpha\alpha}%
$\ might not be negligible in all cases. \ Consider that $M_{\alpha\alpha}$ is
at most of order $\left(  \left\vert \mathbf{U}\right\vert /\upsilon\right)
\left(  r/L\right)  ^{2}$ relative to $E_{\alpha\alpha}$. \ Although $\left(
r/L\right)  ^{2}\ll1$ follows from (\ref{loho1}), $\left\vert \mathbf{U}%
\right\vert /\upsilon$ can be much larger than unity. \ Thus, $M_{\alpha
\alpha}$ cannot be neglected relative to $E_{\alpha\alpha}$ on the basis of
(\ref{loho1}); the same is true for $\frac{1}{2}\left(  U_{n}+U_{n}^{\prime
}\right)  \partial_{X_{n}}\widehat{D}_{\alpha\alpha}$ because it is also
proportional to $\left\vert \mathbf{U}\right\vert /\upsilon$. \ We assume that
the mean flow does not have an abrupt change near the positions of the
anemometers. \ Then, use of (\ref{inertial2}) and (\ref{viscous}) shows that
$r^{2}G_{\alpha,1}G_{\alpha,1}$\ is of order $\left(  r/L\right)  ^{4/3}$ and
$\left(  \ell/L\right)  ^{4/3}$ relative to $\widehat{D}_{\alpha\alpha}$\ in
the inertial and viscous ranges, respectively. \ However, it is not clear on
this basis that\ we can neglect $M_{\alpha\alpha}$ relative to $\frac{1}%
{2}\left(  U_{n}+U_{n}^{\prime}\right)  \partial_{X_{n}}\widehat{D}%
_{\alpha\alpha}$ because what is needed in (\ref{effreyn}) is the streamwise
rate of change, i.e., $\left(  U_{n}+U_{n}^{\prime}\right)  \partial_{X_{n}}$,
operating on both $r^{2}G_{\alpha,1}G_{\alpha,1}$ and $\widehat{D}%
_{\alpha\alpha}$. \ Consequently, we will not further simplify (\ref{effhighR}).

\qquad Now consider the term $\partial_{t}D_{ij}$\ in (\ref{exactave}).
\ Recall that the positions of the anemometers, namely $\mathbf{x}$ and
$\mathbf{x}^{\prime}$, are held fixed for the time-derivative operation
$\partial_{t}$. \ Thus, the meaning of $\partial_{t}D_{ij}$ is the time rate
of change of $D_{ij}$ in the anemometer's rest frame.\ \ The sum of
$\partial_{t}D_{\alpha\alpha}$ and $\frac{1}{2}\left(  U_{n}+U_{n}^{\prime
}\right)  \partial_{X_{n}}D_{\alpha\alpha}$ [see (\ref{effhighR})] is the time
rate of change of $D_{\alpha\alpha}$ in the reference frame moving with
velocity $\left(  \mathbf{U}+\mathbf{U}^{\prime}\right)  /2$; that is, moving
with\ the fluid in the sense of moving with the local and momentary
ensemble-averaged velocity. \ Now (\ref{exactave}) is exact and therefore
describes cases that include rapid changes of mean conditions in the rest
frame of the anemometers. \ However, assume that the experimenter has chosen a
case for which mean conditions are nearly constant in the anemometer's rest
frame; examples include fixed anemometer positions in a wind tunnel, pipe, or
jet for constant mean flow, or freely decaying DNS. \ From the Reynolds
decomposition (\ref{ReynD2}) we have
\begin{equation}
\partial_{t}D_{\alpha\alpha}=r^{2}\partial_{t}\left(  G_{\alpha,1}G_{\alpha
,1}\right)  +\partial_{t}\widehat{D}_{\alpha\alpha}. \label{dalpalpdt}%
\end{equation}
For example, consider the case of turbulence that is freely decaying in the
anemometer's rest frame, or freely decaying DNS. \ In this case,
$r^{2}\partial_{t}\left(  G_{\alpha,1}G_{\alpha,1}\right)  $ is at most of
order $\left(  r/L\right)  ^{2}$ relative to $E_{\alpha\alpha}$, whereas for
the inertial and viscous ranges $\partial_{t}\widehat{D}_{\alpha\alpha}$ is at
most of orders $\left(  r/L\right)  ^{2/3}$ and $\left(  \ell/L\right)
^{2/3}\left(  r/L\right)  ^{2}$ relative to $E_{\alpha\alpha}$. \ For this
case, \ $\partial_{t}D_{\alpha\alpha}$ can be neglected in (\ref{exactave}).
\ More generally, $\partial_{t}D_{\alpha\alpha}$ is negligible because the
experimenter chooses not to move the anemometers rapidly through positions
where mean conditions differ greatly. \ Given the opposite choice,
$\partial_{t}D_{\alpha\alpha}$ would not be negligible; it would be of order
$\left(  r/L\right)  ^{2}\left(  \left|  \mathbf{V}\right|  /\upsilon\right)
$ relative to $E_{\alpha\alpha}$, where $\left|  \mathbf{V}\right|  $ is the
speed of the anemometers relative to the large-scale inhomogeneous structures
of the mean flow. \ Although $\left(  r/L\right)  ^{2}$ is small compared with
unity, $\left(  \left|  \mathbf{V}\right|  /\upsilon\right)  $ can be made
large by increasing the speed of the anemometers relative to the mean-flow
structure. \ Thus, the term $\partial_{t}D_{\alpha\alpha}$ cannot be neglected
from (\ref{exactave}) solely on the basis of (\ref{loho1}) for the same reason
that applies to $\partial_{X_{n}}F_{ijn}$. \ We do neglect\ $\partial
_{t}D_{\alpha\alpha}$ on the basis of the choice mentioned above.

\qquad Reconsider the term $M_{\alpha\alpha}$ in (\ref{effreyn}) together with
$r^{2}\partial_{t}\left(  G_{\alpha,1}G_{\alpha,1}\right)  $, which appears in
(\ref{dalpalpdt}). \ Their sum, i.e., $\left[  \partial_{t}+\frac{1}{2}\left(
U_{n}+U_{n}^{\prime}\right)  \partial_{X_{n}}\right]  r^{2}\left(
G_{\alpha,1}G_{\alpha,1}\right)  $, is the temporal rate of change following
the mean flow of $r^{2}\left(  G_{\alpha,1}G_{\alpha,1}\right)  $. \ This
might not be negligible for some flows, such as a contraction in a wind tunnel
or an expanding round jet, even though $\partial_{t}\left(  G_{\alpha
,1}G_{\alpha,1}\right)  $ might be zero.\ \ This helps illustrate that
$M_{\alpha\alpha}$ might not be negligible.

\qquad Now consider the term proportional to $\nu$ in (\ref{exactave}). \ The
term $\frac{1}{4}\partial_{X_{n}}\partial_{X_{n}}D_{ij}$ is of order $\left(
r/L\right)  ^{2}$ relative to $\partial_{r_{n}}\partial_{r_{n}}D_{ij} $, and
is negligible. \ The Laplacian operating on (\ref{ReynD2}) gives
$\partial_{r_{n}}\partial_{r_{n}}D_{ij}\simeq6G_{i,1}G_{j,1}+\partial_{r_{n}%
}\partial_{r_{n}}\widehat{D}_{ij}$. \ Now, $2\nu\left(  6G_{i,1}%
G_{j,1}\right)  $ is at most of order $2\nu\left(  \upsilon/L\right)  ^{2}$,
which is of orders $\left(  r/L\right)  ^{4/3}$ and $\left(  \ell/L\right)
^{4/3}$ relative to (\ref{arbproD2lapinert}) and (\ref{arbproD2lapvisc}),
respectively.\ \ Therefore, (\ref{loho1}) and (\ref{arbproD2lapinert}) and
(\ref{arbproD2lapvisc}) give $\breve{a}_{i}\breve{a}_{j}\left(  \partial
_{r_{n}}\partial_{r_{n}}D_{ij}+\frac{1}{4}\partial_{X_{n}}\partial_{X_{n}%
}D_{ij}\right)  \simeq\breve{a}_{i}\breve{a}_{j}\partial_{r_{n}}%
\partial_{r_{n}}\widehat{D}_{ij}$.

\qquad In the Reynolds decomposition of $E_{\alpha\alpha}$ the terms that
depend on mean velocity are of the order of an inverse Reynolds number
$\left(  \upsilon L/\nu\right)  ^{-1}\ll1$ relative to $\left\langle
\varepsilon\right\rangle _{E}$. \ Thus, (\ref{ediagonal}) gives $E_{\alpha
\alpha}\simeq\widehat{E}_{\alpha\alpha}$. \ By the same method, use of the
definition (\ref{strain}) of $\varepsilon$\ gives $\left\langle \varepsilon
\right\rangle _{E}\simeq\left\langle \widehat{\varepsilon}\right\rangle _{E}$.
\ That is, the mean velocity produces negligible viscous dissipation.

\qquad In the average of (\ref{eii}), consider the term $2\nu\partial_{X_{n}%
}\partial_{X_{n}}\left\langle p+p^{\prime}\right\rangle _{E}$, which also
appears in (\ref{W}). \ Excluding the case of nearby bodies in the flow that
can cause sharp spatial variation of pressure, the mean pressure gradient
scales with $\upsilon$ and $L$. \ Then, the term $2\nu\partial_{X_{n}}%
\partial_{X_{n}}\left\langle p+p^{\prime}\right\rangle _{E}$\ is of order
$\left(  \upsilon L/\nu\right)  ^{-1}$ relative to $\left\langle
\varepsilon\right\rangle _{E}$, and is thus negligible. \ The Taylor series
expansion (\ref{Taylor seriesdissip}) shows that $\left\langle \varepsilon
+\varepsilon^{\prime}\right\rangle _{E}\simeq2\left\langle \varepsilon
\right\rangle _{E}$, where the neglected terms are at most of order $\left(
r/L\right)  ^{2}$ relative to $\left\langle \varepsilon\right\rangle _{E}$ and
are therefore negligible on the basis of (\ref{loho1}). \ Then, the average of
(\ref{eii}) gives the trace: $\left\langle e_{ii}\right\rangle _{E}\equiv
E_{ii}\simeq\widehat{E}_{ii}\simeq4\left\langle \varepsilon\right\rangle
_{E}\simeq4\left\langle \widehat{\varepsilon}\right\rangle _{E}$.

\qquad Finally, consider the Reynolds decomposition of $T_{ij}$. \ Denote the
mean pressure gradient at point $\mathbf{X}$\ by $\Pi_{n}\equiv\partial
_{X_{n}}\left\langle p\left(  \mathbf{X},t\right)  \right\rangle _{E}$. \ The
Reynolds decomposition of the term $-2\left\langle \left(  p-p^{\prime
}\right)  \left(  s_{ij}-s_{ij}^{\prime}\right)  \right\rangle _{E}$ in
$T_{ij} $ [see (\ref{tau2})] gives a mean-gradients term that is approximated
by $-r_{q}r_{n}\Pi_{n}\partial_{Xq}\left(  G_{j,i}+G_{i,j}\right)  =-r^{2}%
\Pi_{1}\partial_{X_{1}}\left(  G_{j,i}+G_{i,j}\right)  $. \ Recall that
$\Pi_{n} $ scales with $\upsilon$ and $L$. Then, $-r^{2}\Pi_{1}\partial
_{X_{1}}\left(  G_{j,i}+G_{i,j}\right)  $ is of order $\left(  r/L\right)
^{2}$ relative to $E_{\alpha\alpha}$, such that this term is negligible in
(\ref{exactave}). \ Using (\ref{Tinertial1}) and (\ref{Tviscous}) for the
diagonal components of $-2\left\langle \left(  \widehat{p}-\widehat{p}%
^{\prime}\right)  \left(  \widehat{s}_{ij}-\widehat{s}_{ij}^{\prime}\right)
\right\rangle _{E}$ this term is seen to be negligible compared to
$E_{\alpha\alpha}$ for $r$ within the inertial range through the viscous
range. \ The Reynolds decomposition gives\ $\partial_{X_{i}}\left\langle
\left(  p-p^{\prime}\right)  \left(  u_{j}-u_{j}^{\prime}\right)
\right\rangle _{E}\simeq r_{q}r_{n}\partial_{X_{i}}\left(  \Pi_{n}%
G_{j,q}\right)  +\partial_{X_{i}}\left\langle \left(  \widehat{p}-\widehat
{p}^{\prime}\right)  \left(  \widehat{u}_{j}-\widehat{u}_{j}^{\prime}\right)
\right\rangle _{E}$. \ Since $\Pi_{n}$ scales with $\upsilon$ and $L$, the
term $r_{q}r_{n}\partial_{X_{i}}\left(  \Pi_{n}G_{j,q}\right)  =r^{2}%
\partial_{X_{i}}\left(  \Pi_{1}G_{j,1}\right)  $ is at most of order $\left(
r/L\right)  ^{2}$ relative to $E_{\alpha\alpha}$, and this term is therefore
negligible in (\ref{exactave}). \ On the basis of (\ref{Tterm}) and the
neglect of $\ -2\left\langle \left(  \widehat{p}-\widehat{p}^{\prime}\right)
\left(  \widehat{s}_{\alpha\alpha}-\widehat{s}_{\alpha\alpha}^{\prime}\right)
\right\rangle _{E}$, we also neglect $\partial_{X_{\alpha}}\left\langle
\left(  \widehat{p}-\widehat{p}^{\prime}\right)  \left(  \widehat{u}_{\alpha
}-\widehat{u}_{\alpha}^{\prime}\right)  \right\rangle _{E}$. \ Taken together,
these approximations show that $T_{\alpha\alpha}$ is negligible in
(\ref{exactave}). \ On the other hand, mean pressure gradient can be large in
the presence of bodies in the flow; a contraction of a wind tunnel is an
example. \ Thus, like $\partial_{X_{n}}F_{ijn}$, terms containing the mean
pressure gradient cannot be neglected on the basis of (\ref{loho1}) alone.
\ In effect, we have assumed that there are no bodies strongly affecting the
local turbulent flow. \ For this case, $T_{\alpha\alpha}$ is negligible in the
diagonal elements of (\ref{exactave}).

\qquad The results of the above scale analysis are summarized in the following
three sections.

\bigskip

\begin{center}
\textbf{8. \ APPROXIMATE EQUATIONS\bigskip}

\textbf{8.1 \ Ensemble Average: \ Approximate Equations\bigskip}
\end{center}

\qquad Given the experimental case discussed above and quantified in Appendix
B, the diagonal elements of (\ref{exactave}) projected in arbitrary directions
$\mathbf{\breve{a}}$ give the approximate equation
\begin{equation}
\breve{a}_{i}\breve{a}_{j}\left[  \frac{1}{2}\left(  U_{n}+U_{n}^{\prime
}\right)  \partial_{X_{n}}D_{ij}+\partial_{r_{n}}\widehat{D}_{ijn}%
=2\nu\partial_{r_{n}}\partial_{r_{n}}\widehat{D}_{ij}-\widehat{E}_{ij}\right]
.\label{approxave}%
\end{equation}
As examples, the direction $\mathbf{\breve{a}}$ can be chosen to be in the
direction of some large-scale flow symmetry, such as streamwise or cross
stream, etc., or in a direction defined by the separation of anemometers, such
as $\mathbf{r}$ or perpendicular to $\mathbf{r}$. \ The appearance
of\ $D_{ij}$, rather than $\widehat{D}_{ij}$, in the left-most term in
(\ref{approxave}) indicates that both terms in (\ref{effreyn}) are included.
$\ $The trace of (\ref{exactave}) becomes
\begin{equation}
\frac{1}{2}\left(  U_{n}+U_{n}^{\prime}\right)  \partial_{X_{n}}%
D_{ii}+\partial_{r_{n}}\widehat{D}_{iin}=2\nu\partial_{r_{n}}\partial_{r_{n}%
}\widehat{D}_{ii}-4\left\langle \varepsilon\right\rangle _{E}%
.\label{approxtrace}%
\end{equation}
As shown above, derivation of (\ref{approxave}) and (\ref{approxtrace}) from
the exact equation (\ref{exactave}) requires more than just (\ref{loho1}). \ A
further requirement is that the experimenter avoids cases having large spatial
and temporal variation of the mean flow. \ Of course, that choice improves the
accuracy of local homogeneity for fixed values of $\left[  \max\left(
r,\ell\right)  /L\right]  $. \ Additional requirements are approximations
(\ref{gammapprox}) and (\ref{Tterm}), and that the inverse Reynolds number
$\left(  \upsilon L/\nu\right)  ^{-1}$\ is very small. \ In general, those
conditions are typical of an experimental situation that is sought for the
study of the universality of turbulence statistics at small scales. \ Most
experiments use Taylor's hypothesis to estimate spatial statistics from
temporal statistics, for which purpose $\left\vert \mathbf{U}\right\vert
/\upsilon$ must be large. \ For this reason, the left-most term is not
neglected in (\ref{approxave}), nor in (\ref{approxtrace}).

\qquad Of course, (\ref{approxave}) contains no information about the
off-diagonal elements of (\ref{exactave}). \ We cannot evaluate those
off-diagonal elements because we lack the necessary data. \ Clearly, DNS or a
very complete experiment (e.g., as in Su and Dahm, 1996)) could be used to
quantify those off-diagonal elements. \ The off-diagonal elements of
(\ref{exactave}) describe quantities that approach zero as local isotropy
becomes accurate.

\bigskip\pagebreak

\begin{center}
\textbf{8.2 \ Temporal Average: \ Approximate Equations\bigskip}
\end{center}

\qquad\ \ Using (\ref{tempavederiv}), we noted the case for which
$\left\langle \partial_{t}d_{ij}\right\rangle _{T}$ can be made as small as
desired by use of a long averaging duration. \ This case is typical of
experimental work for which the temporal average is also typical. \ Assume
that this is the case such that in (\ref{tempexact2}) $\breve{a}_{i}\breve
{a}_{j}\left\langle \partial_{t}d_{ij}\right\rangle _{T}$ can be neglected,
and, in the case of (\ref{tempexactrace}) that $\left\langle \partial
_{t}d_{ii}\right\rangle _{T}$ can be neglected. \ On the other hand, recall
from (\ref{tempavederiv}) that it is easy to evaluate $\breve{a}_{i}\breve
{a}_{j}\left\langle \partial_{t}d_{ij}\right\rangle _{T}$\ from by use of
experimental data. \ The Reynolds decomposition and the approximations that
lead from (\ref{exactave}) and (\ref{exactrace}) to (\ref{approxave}) and
(\ref{approxtrace}) also apply to (\ref{tempexact2}) and (\ref{tempexactrace}%
); \ we immediately obtain
\begin{equation}
\breve{a}_{i}\breve{a}_{j}\left[  \frac{1}{2}\left(  U_{n}+U_{n}^{\prime
}\right)  \partial_{X_{n}}\left\langle d_{ij}\right\rangle _{T}+\partial
_{r_{n}}\left\langle \widehat{d}_{ijn}\right\rangle _{T}=2\nu\partial_{r_{n}%
}\partial_{r_{n}}\left\langle \widehat{d}_{ij}\right\rangle _{T}-\left\langle
\widehat{e}_{ij}\right\rangle _{T}\right]  ,\label{tempapproxave}%
\end{equation}%
\begin{equation}
\frac{1}{2}\left(  U_{n}+U_{n}^{\prime}\right)  \partial_{X_{n}}\left\langle
d_{ii}\right\rangle _{T}+\partial_{r_{n}}\left\langle \widehat{d}%
_{iin}\right\rangle _{T}=2\nu\partial_{r_{n}}\partial_{r_{n}}\left\langle
\widehat{d}_{ii}\right\rangle _{T}-4\left\langle \widehat{\varepsilon
}\right\rangle _{T},\label{aatempapprotrace2}%
\end{equation}
where, as before, the caret over the averaged quantity means that the quantity
is calculated from fluctuations. \ These equations relate the statistics that
experimenters (e.g., Antonia, Chambers, and Browne, 1983; Chambers and
Antonia, 1984; Danaila \textit{et al.}, 1999 a,b) calculate from data. \ As
shown in Sec. 5, the mean quantities, i.e., $U_{n}(\mathbf{x},t_{0}%
,T)\equiv\left\langle u_{n}(\mathbf{x},t)\right\rangle _{T}$, in the
definition of the Reynolds decomposition (\ref{defineReydecomp}) are now time
averages rather than ensemble averages such that $\left\langle \widehat{u}%
_{i}(\mathbf{x},t)\right\rangle _{T}=0$, etc. \ Except for replacing the
ensemble average with the time average, (\ref{tempapproxave}) and
(\ref{aatempapprotrace2}) are the same as (\ref{approxave}) and
(\ref{approxtrace}). \ However, the statistics in (\ref{approxave}) and
(\ref{approxtrace}) can have dependence on $t$, whereas the statistics in
(\ref{tempapproxave}) and (\ref{aatempapprotrace2}) depend on only the time of
the start of the temporal average (i.e., $t_{0}$) and the duration of the
average ($T$); in addition to which the dependence on start time and duration
must be slight because of the neglect of $\left\langle \partial_{t}%
d_{ij}\right\rangle _{T}$.

\bigskip

\begin{center}
\textbf{8.3 \ Spatial Average: \ Approximate Equations\bigskip}
\end{center}

\qquad Now consider spatial averaging. \ Given the approximations that lead
from (\ref{exactave}) and (\ref{exactrace}) to (\ref{approxave}) and
(\ref{approxtrace}), (\ref{spatialaveext2}) and (\ref{spatialavetrace})
become
\begin{equation}
\breve{a}_{i}\breve{a}_{j}\left[  \partial_{t}\left\langle \widehat{d}%
_{ij}\right\rangle _{\mathbb{R}}+\frac{S}{2V}\oint_{\mathbf{X}_{n}}\left(
U_{n}+U_{n}^{\prime}\right)  d_{ij}+\partial_{r_{n}}\left\langle \widehat
{d}_{ijn}\right\rangle _{\mathbb{R}}=2\nu\partial_{r_{n}}\partial_{r_{n}%
}\left\langle \widehat{d}_{ij}\right\rangle _{\mathbb{R}}-\left\langle
\widehat{e}_{ij}\right\rangle _{\mathbb{R}}\right]  , \label{approxspace}%
\end{equation}
\begin{equation}
\partial_{t}\left\langle \widehat{d}_{ii}\right\rangle _{\mathbb{R}}+\frac
{S}{2V}\oint_{\mathbf{X}_{n}}\left(  U_{n}+U_{n}^{\prime}\right)
d_{ii}+\partial_{r_{n}}\left\langle \widehat{d}_{iin}\right\rangle
_{\mathbb{R}}=2\nu\partial_{r_{n}}\partial_{r_{n}}\left\langle \widehat
{d}_{ii}\right\rangle _{\mathbb{R}}-4\left\langle \widehat{\varepsilon
}\right\rangle _{\mathbb{R}}. \label{approxspacetrace}%
\end{equation}
As shown in Sec. 5, the mean quantities, $U_{n}(t)\equiv\left\langle
u_{n}(\mathbf{x},t)\right\rangle _{\mathbb{R}}$, in the definition of the
Reynolds decomposition (\ref{defineReydecomp}) are now space averages rather
than ensemble averages such that $\left\langle \widehat{u}_{i}(\mathbf{x}%
,t)\right\rangle _{\mathbb{R}}=0$, etc. \ As in the previous case, the caret
above a quantity designates that it is calculated from velocity fluctuations.
\ The time-derivative terms $\partial_{t}\left\langle \widehat{d}%
_{ij}\right\rangle _{\mathbb{R}}$\ and $\partial_{t}\left\langle \widehat
{d}_{ii}\right\rangle _{\mathbb{R}}$ have been retained in (\ref{approxspace})
and (\ref{approxspacetrace}) because they are more significant than the
advective term for the case of freely decaying DNS. \ Another example is the
forced DNS flow of Borue and Orszag (1996), because it exhibits temporal
variation of total mean-squared vorticity by a factor of 2. \ It seems prudent
to retain the time derivatives. \ For DNS data, the advective term in both
(\ref{approxspace}) and (\ref{approxspacetrace}) is seldom important.
\ Consider the DNS flow of Borue and Orszag (1996), for which $\left\vert
\mathbf{U}\right\vert /\upsilon$ was at most about 2. \ Then, on the basis of
the scale analysis [see below (\ref{effreyn})], the advective term is
negligible on the basis of (\ref{loho1}). \ In (\ref{approxspace}) there is no
information on the off-diagonal components because the approximations apply
only to the diagonal components.

\qquad Also, (\ref{approxspace}) and (\ref{approxspacetrace})\ become
\[
\breve{a}_{i}\breve{a}_{j}\left[  \partial_{t}\left\langle \widehat{d}%
_{ij}\right\rangle _{\mathbb{R}}+\partial_{r_{n}}\left\langle \widehat
{d}_{ijn}\right\rangle _{\mathbb{R}}=2\nu\partial_{r_{n}}\partial_{r_{n}%
}\left\langle \widehat{d}_{ij}\right\rangle _{\mathbb{R}}-\left\langle
\widehat{e}_{ij}\right\rangle _{\mathbb{R}}\right]  ,
\]
\begin{equation}
\partial_{t}\left\langle \widehat{d}_{ii}\right\rangle _{\mathbb{R}}%
+\partial_{r_{n}}\left\langle \widehat{d}_{iin}\right\rangle _{\mathbb{R}%
}=2\nu\partial_{r_{n}}\partial_{r_{n}}\left\langle \widehat{d}_{ii}%
\right\rangle _{\mathbb{R}}-4\left\langle \widehat{\varepsilon}\right\rangle
_{\mathbb{R}}. \label{spacapproxlasttrac}%
\end{equation}

\bigskip

\begin{center}
\textbf{9. \ DISCUSSION\bigskip}
\end{center}

\qquad Given data for which local homogeneity and/or local isotropy are
approximate, it seems that (\ref{exactrace}) is closer to that asymptotic case
than is (\ref{exactave}), and therefore, that data for the trace $D_{iin}$
will more accurately show the asymptotic inertial-range power law than does
$D_{111}$. \ The reason is as follows. \ For the approach toward local
isotropy in homogeneous turbulence, the anisotropy quantified by nonzero
values of $T_{ij}$ is balanced by that from the term $\partial_{r_{n}}D_{ijn}$
in (\ref{exactave}) (Hill, 1997). \ The trace of $T_{ij}$ vanishes exactly for
the homogeneous case because $\partial_{X_{i}}\left\langle \left(
p-p^{\prime}\right)  \left(  u_{i}-u_{i}^{\prime}\right)  \right\rangle
_{E}=0$ for homogeneous turbulence and because $-2\left\langle \left(
p-p^{\prime}\right)  \left(  s_{ii}-s_{ii}^{\prime}\right)  \right\rangle
_{E}=0$ on the basis of incompressibility ($s_{ii}=0$). \ Then, $\partial
_{r_{n}}D_{iin}$ must balance less anisotropy in (\ref{exactrace}) than does
$\partial_{r_{n}}D_{ijn}$ in (\ref{exactave}). \ For inhomogeneous turbulence,
the nonvanishing part of the trace,\ namely $T_{ii}=2\partial_{X_{i}%
}\left\langle \left(  p-p^{\prime}\right)  \left(  u_{i}-u_{i}^{\prime
}\right)  \right\rangle _{E}$, is expected to approach zero rapidly as $r$
decreases for two reasons. \ First, $\left\langle \left(  p-p^{\prime}\right)
\left(  u_{i}-u_{i}^{\prime}\right)  \right\rangle _{E}$ vanishes on the basis
of local isotropy. \ Second, the operator $\partial_{X_{i}}$ causes
$\partial_{X_{i}}\left\langle \left(  p-p^{\prime}\right)  \left(  u_{i}%
-u_{i}^{\prime}\right)  \right\rangle _{E}$\ to vanish on the basis of local
homogeneity. \ The right-most two terms in (\ref{W}) contain the operator
$\partial_{X_{n}}\partial_{X_{n}}$, which causes these terms in $W$ to vanish
rapidly on the basis of local homogeneity. \ Thus, all terms in $W$ are
negligible for locally homogeneous turbulence. \ By performing the trace it
appears that anisotropy has been significantly reduced in (\ref{exactrace})
relative to in (\ref{exactave}). \ It follows that the trace, $\partial
_{r_{n}}D_{iin}$, is affected less by anisotropy than is $\partial_{r_{n}%
}D_{ijn}$, and therefore, that $D_{iin}$ is less affected by anisotropy than
is $D_{ijn}$. \ This hypothesis should be checked by comparison with DNS.
\ Evaluation of all terms in (\ref{exactrace}) and (\ref{exactave}) are the
basis for such an investigation. \ We therefore expect that inertial-range
power-law scaling would be more evident in $D_{ii1}$ than in $D_{111}$. \ Of
course, performing the trace requires that all three components of velocity be
measured at both $\mathbf{x}$ and $\mathbf{x}^{\prime}$.

\qquad To determine scaling properties of the third-order structure function,
past theory has used the isotropic-tensor formula to produce a differential
equation having the operator $\partial_{r}$ and integration of that equation
(as done in Sec. 4.3). \ However, one can use an equation like
(\ref{tempexactrace}) without an assumption about the symmetry properties
(e.g., isotropic) of the structure functions by means of the sphere average in
$\mathbf{r}$-space, as implemented in Sec. 4.2. \ Evaluating resultant terms
in the $\mathbf{r}$-space sphere-averaged equation implies a tedious
experimental procedure if wire anemometers are used. \ On the other hand, both
DNS and the experimental method of Su and Dahm (1996) are suited to such
evaluation. \ In effect, the $\mathbf{r}$-space sphere average solves the
equation by producing the orientation-averaged third-order structure function.
\ It would seem that the orientation average mitigates anisotropy
effects.\ \ Thus, the orientation average of the trace of the third-order
structure function, namely, $\oint_{\mathbf{r}_{n}}D_{ii1}$, is expected to
best exhibit properties of locally isotropic turbulence, such as the
inertial-range power law with the $4/3$ coefficient that appears in
(\ref{simpleDNS}).

\qquad Lindborg (1999) estimates the contribution of $\frac{1}{2}\left(
U_{n}+U_{n}^{\prime}\right)  \partial_{X_{n}}D_{\alpha\alpha}$ (for the case
$U_{n}=U_{n}^{\prime}$) to experimental measurements of $\left\langle
\widehat{d}_{111}\right\rangle _{T}$ for grid, jet, and wake turbulence of
moderate Reynolds number, and Danaila \textit{et al.} (1999 a,b) do so for
grid turbulence at $R_{\lambda}=66$, $99$, and $448$. \ They show that the
term $\frac{1}{2}\left(  U_{n}+U_{n}^{\prime}\right)  \partial_{X_{n}%
}D_{\alpha\alpha}$ accounts for much of the observed deviation of the data
from Kolmogorov's equation; Kolmogorov's equation is $\left\langle \widehat
{d}_{111}\right\rangle _{T}=6\nu\partial_{r}\left\langle \widehat{d}%
_{11}\right\rangle _{T}-\frac{4}{5}\left\langle \widehat{\varepsilon
}\right\rangle _{T}r$. \ In the case of Danaila \textit{et al.} (1999a), one
must keep in mind that their estimation method reduces their equation to
$2\left\langle \widehat{u_{1}}^{2}\right\rangle =\underset{r\rightarrow\infty
}{\lim}\widehat{D}_{11}$ in the energy-containing range such that the balance
of the equation is not tested in the energy-containing range.

\bigskip

\begin{center}
\textbf{10. \ CONCLUSION\bigskip}
\end{center}

\qquad The mathematical method of deriving exact structure-function equations
from the Navier-Stokes equation is developed. \ The basic tools are the change
of variables (\ref{change}) and the derivative identities (\ref{derivs}) and
(\ref{identderivs}) and algebra. \ Manipulations are performed to the greatest
extent possible (in Sec. 2) before an average is performed. \ Then, exactly
defined ensemble, time, and spatial averages are used. \ DNS makes study of
exact structure-function equations practical. \ Also, experimental methods
exist (Su and Dahm, 1995) that can completely evaluate terms in the exact
structure-function equations. \ Exact incompressibility relationships, such as
(\ref{avesecdinc1}) and (\ref{avesecdinc2}), are obtained. \ Following from
the discussion in Sec. 9, the exact incompressibility relationship
(\ref{avesecdinc1}) will have a nonzero value at small $r$ because of
large-scale structures in the flow. \ At small $r$, (\ref{avesecdinc2}) is
approximately the second derivative with respect to measurement location of
the velocity covariance, and therefore clearly depends on flow inhomogeneity.

\qquad That the exact structure-function equations are an advance can be seen
from previous work. \ It is no longer necessary to derive individual terms
that describe effects of inhomogeneity that are missing from equations valid
only for homogeneous turbulence, such as was done by Lindborg (1999). \ All
such terms are now known. \ Sreenivasan and Dhruva (1998) note that one could
determine scaling exponents with greater confidence if one has a theory that
exhibits not only the asymptotic power law but also the trend toward the power
law, and that without such a theory the method of computing local slopes is a
\textquotedblleft misplaced delusion.\textquotedblright\ \ The exact equations
given here are the required theory for the third-order structure function,
given that data must be used to evaluate the equations in a manner analogous
to previous evaluations in Antonia, Chambers, and Browne (1983), Chambers and
Antonia (1984), Lindborg (1999),\ Danaila \textit{et al.} (1999 a,b), and
Antonia \textit{et al.} (2000). \ The exact dynamical equations obtained here
are useful for studies of the approach toward local homogeneity as well as to
local isotropy. \ Toward that end, a scale analysis is given in Sec. 7.4,
which leads to the approximate equations in Sec. 8. \ The exact equations
provide insight into the time-derivative terms, as discussed in Sec. 6.

\bigskip

\begin{center}
\textbf{11. \ ACKNOWLEDGMENT\bigskip}
\end{center}

\qquad The author thanks Dr. Mikhail Charnotskii and Dr. Eric Lindborg for
helpful comments.

\bigskip

\begin{center}
\textbf{12. \ REFERENCES}\bigskip
\end{center}

\noindent Alvelius, K., and A. V. Johansson, 2000. LES computations and
comparison with

\qquad Kolmogorov theory for two-point pressure-velocity correlations and structure

\qquad functions for globally anisotropic turbulence. \textit{J. Fluid Mech}. 403:23-36.

\noindent Anselmet, F., E. J. Gagne, and E. J. Hopfinger, 1984. High-order
velocity structure functions

\qquad in turbulent shear flows. \textit{J. Fluid Mech}. 140:63-89.

\noindent Antonia, R. A., A. J. Chambers, and L. W. B. Browne, 1983. Relations
between structure

\qquad functions of velocity and temperature in a turbulent jet.
\textit{Experiments in Fluids}

\qquad1:213-219.

\noindent Antonia, R. A., T. Zhou, L. Danaila, and F. Anselmet, 2000.
Streamwise inhomogeneity of

\qquad decaying grid turbulence. \textit{Phys. Fluids} 12:3086-3089.

\noindent Batchelor, G. K., 1947. Kolmogoroff's theory of locally isotropic
turbulence. \textit{Proc.}

\qquad\textit{Cambridge. Philos. Soc}. 43:533-559.

\noindent Batchelor, G. K., 1956.\ \textit{The Theory of Homogeneous
Turbulence}. Cambridge University

\qquad Press, 195 pp.

\noindent Belin, F., J. Maurer, P. Tabeling, and H. Willaime, 1997. Velocity
gradient distributions in

\qquad fully developed turbulence: \ An experimental study. \textit{Phys.
Fluids} 9:3843-3850.

\noindent Boratav, O. N., and R. B. Pelz, 1997. Structures and structure
functions in the inertial

\qquad range of turbulence. \textit{Phys. Fluids} 9:1400-1415.

\noindent Borue, V., and S. A. Orszag, 1996. Numerical study of
three-dimensional Kolmogorov flow

\qquad at high Reynolds numbers. \textit{J. Fluid Mech}. 306:293-323.

\noindent Chambers, A. J., and R. A. Antonia, 1984. Atmospheric estimates of
power-law exponents

\qquad\ $\mu$\ and $\mu_{\theta}$. \textit{Bound.-Layer Meteorol.} 28:343-52.

\noindent Danaila, L., F. Anselmet, T. Zhou, and\ R. A Antonia, 1999a. A
generalization of Yaglom's

\qquad equation which accounts for the large-scale forcing in heated decaying turbulence.

\qquad J\textit{. Fluid Mech.} 391:359-372.

\noindent Danaila, L., P. Le Gal, F. Anselmet, F. Plaza, and J. F. Pinton,
1999b. Some new features

\qquad of the passive scalar mixing in a turbulent flow. \textit{Phys. Fluids} 11:636-646.

\noindent de Bruyn Kops, S. M., and J. J. Riley, 1998. Direct numerical
simulation of laboratory

\qquad experiments in isotropic turbulence. \textit{Phys. Fluids} 10:2125-2127.

\noindent Frisch, U. 1995. \textit{Turbulence, The Legacy of A. N.
Kolmogorov}. Cambridge University Press,

\qquad288 pp.

\noindent Hill, R. J., and J. M. Wilczak, 2001. Fourth-order velocity
statistics. \textit{Fluid Dyn. Res.}

\qquad28:1-22.

\noindent Hill, R. J., 1997. Applicability of Kolmogorov's and Monin's
equations of turbulence.

\qquad\textit{J.~Fluid Mech.} 353:67-81.

\noindent Hill, R. J., 2001. Equations relating structure functions of all
orders. \textit{J. Fluid Mech.}

\qquad434:379-388.

\noindent Kaimal, J. C., J. C. Wyngaard, D. A. Haugen, O. R. Cote, and Y
Izumi, 1976. Turbulence

\qquad structure in the convective boundary layer. \textit{J. Atmos. Sci.} 33:2152-2169.

\noindent Kolmogorov, A. N., 1941a. Dissipation of energy in locally isotropic
turbulence. \textit{Dokl. Akad.}

\qquad\textit{Nauk SSSR} 32:16-18.

\noindent Kolmogorov, A. N., 1941b. The local structure of turbulence in
incompressible viscous fluid

\qquad for very large Reynolds numbers. \textit{Dokl. Akad. Nauk SSSR} 30:301-305.

\noindent Kolmogorov, A. N., 1962.\ A refinement of previous hypotheses
concerning the local structure

\qquad of turbulence in a viscous incompressible fluid at high Reynolds
number. \textit{J.~Fluid}

\qquad\textit{Mech.} 13:82-85.

\noindent Lindborg, E., 1996. A note on Kolmogorov's third-order
structure-function law, the local

\qquad isotropy hypothesis and the pressure-velocity correlation.
\textit{J.~Fluid Mech.} 326:343-

\qquad356.

\noindent Lindborg, E., 1999. Correction to the four-fifths law due to
variations of the dissipation.

\qquad\textit{Phys. Fluids} 11:510-512.

\noindent Monin, A. S., 1959. The theory of locally isotropic turbulence.
\textit{Dokl. Akad. Nauk. SSSR}

\qquad125:515-518.

\noindent Monin, A. S., and A. Yaglom, 1975. \textit{Statistical Fluid
Mechanics:Mechanics of Turbulence},

\qquad Vol. 2, The MIT Press, Cambridge, MA, 874 pp.

\noindent Mydlarski, L., and Z. Warhaft, 1996. On the onset of
high-Reynolds-number grid-generated

\qquad wind tunnel turbulence. \textit{J. Fluid Mech.} 320:331-368.

\noindent Mydlarski, L., and Z. Warhaft, 1998. Passive scalar statistics in
high-Peclet-number grid

\qquad generated turbulence. \textit{J. Fluid Mech.} 358:135-175.

\noindent Novikov, E. A., 1965. Functionals and the random-force method in
turbulence theory. \textit{Sov.}

\qquad\textit{Phys. JETP} 20:1290-1294.

\noindent Obukhov, A. M., 1949. The structure of the temperature field in a
turbulent flow. \textit{Izv.}

\qquad\textit{Akad. Nauk. SSSR, Ser. Geogr. i Geofiz.} 13:58-69.

\noindent Obukhov, A. M., 1962. Some specific features of atmospheric
turbulence. \textit{J. Fluid Mech.}

\qquad13:77-81.

\noindent Pearson, B. R., P.-A. Krogstad, W. vande Water, 2002. Measurements
of the turbulent energy

\qquad dissipation rate. \textit{Phys. Fluids.} 14:1288-1290.

\noindent Praskovsky, A. A., E. B. Gledzer, M. Yu. Karyakin, and Ye Zhow,
1993. The sweeping

\qquad decorrelation hypothesis and energy-inertial interaction in high
Reynolds number

\qquad flows. \textit{J. Fluid Mech.} 248:493-511.

\noindent Rytov, S., Yu. Kravtsov, and V. Tatarskii, 1989. \textit{Principles
of Statistical Radiophysics 4:}

\qquad\textit{Wave Propagation throught Random Media.} Springer-Verlag,
Berlin, 188 pp.

\noindent Sreenivasan, K. R., 1991. On local isotropy of passive scalars in
turbulent shear flows. \textit{Proc.}

\qquad\textit{R. Soc. London. A} 434:165-182.

\noindent Sreenivasan, K. R., 1998. An update on the energy dissipation rate
in isotropic turbulence.

\qquad\textit{Phys. Fluids} 10:528-529.

\noindent Sreenivasan, K. R., and R. A. Antonia, 1997.\ The phenomenology of
small scale turbulence.

\qquad\textit{Annu. Rev. Fluid Mech.} 29:435-472.

\noindent Sreenivasan K. R., and G. Stolovitzky, 1996. Statistical dependence
of inertial range

\qquad properties on large scales in a high-Reynolds-number shear flow.
\textit{Phys. Rev. E}

\qquad77:2218-2221.

\noindent Sreenivasan, K. R., and B. Dhruva, 1998. Is there scaling in high-Reynolds-number

\qquad turbulence? \textit{Prog. Theor. Phys. Suppl.} 130:103-120.

\noindent Su, L. K., and W. J. A. Dahm, 1996. Scalar imaging velocimetry
measurements of the

\qquad velocity gradient tensor field in turbulent flows. I. Experimental
results. \textit{Phys.}

\qquad\textit{Fluids} 8:1883-1906.

\noindent Tennekes, H., and J. L. Lumley, 1972. \textit{A First Course in
Turbulence,} The MIT Press,

\qquad Cambridge, MA, 300 pp.

\noindent Yaglom, A., 1998. New remarks about old ideas of Kolmogorov.
\textit{Adv. Turb.} VII:605-610.

\noindent Zocchi, G., P. Tabeling, J. Maurer, and H. Willaime, 1994.
Measurement of the scaling of

\qquad the dissipation at high Reynolds numbers. \textit{Phys. Rev. E} 50:3693-3700.

\bigskip

\begin{center}
\bigskip\textbf{Appendix A:\ Forced Turbulence}
\end{center}

\qquad The Navier-Stokes equation (\ref{NSE}) and the exact structure-function
equations [e.g., (\ref{exactave})] apply to cases in which the turbulence is
forced at places other than at the points of observation $\mathbf{x}$ and
$\mathbf{x}^{\prime}$, such as grid-generated turbulence, pipe flow, and
boundary layers. \ Also, the Navier-Stokes equation (\ref{NSE}) and
(\ref{exactave}) apply to freely decaying DNS such as that by Boratav and Pelz
(1997) and the simulation of laboratory experiments as in de Bruyn Kops and J.
J. Riley (1998). \ Some DNS employ spatially distributed forces to drive the
turbulence to a steady state. \ The Navier-Stokes equation (\ref{NSE}), and
the exact equations derived from it, do not apply to that case; instead, such
forces must be introduced into (\ref{NSE}) and the resultant additional terms
derived for the exact structure-function equations.

\qquad If a force $f_{i}$ is added to the right-hand side of the Navier-Stokes
equation (\ref{NSE}), then the term to be added to (\ref{exact2}) is simply
$-\tau_{ij}$ defined in (\ref{tau}) with $-\partial_{x_{i}}p$ replaced by
$f_{i}$ and $-\partial_{x_{i}^{\prime}}p^{\prime}$ by $f_{i}^{\prime}$. \ That
is, the added term is
\[
\left(  f_{i}-f_{i}^{\prime}\right)  \left(  u_{j}-u_{j}^{\prime}\right)
+\left(  f_{j}-f_{j}^{\prime}\right)  \left(  u_{i}-u_{i}^{\prime}\right)
\equiv\text{$\Phi$}_{ij},
\]
and the average of this expression must appear in our subsequent
structure-function equations. \ Consider the case of the deterministic force,
$f_{i}=\delta_{i2}F\cos\left(  k_{f}x_{1}\right)  $, that was used in the DNS
in Borue and Orszag (1996), where we use subscripts $2$ and $1$ to denote
their $y$ and $x$ directions, respectively, and $\delta_{ij}$ is the Kronecker
delta. \ Use the identity $\cos\left(  k_{f}x_{1}\right)  -\cos\left(
k_{f}x_{1}^{\prime}\right)  =2\cos\left(  k_{f}r_{1}/2\right)  \cos\left(
k_{f}X_{1}\right)  $. \ The ensemble and temporal averages of $\Phi_{ij}$ are
$2F\cos\left(  k_{f}r_{1}/2\right)  \cos\left(  k_{f}X_{1}\right)  \left[
\delta_{i2}\left(  U_{j}-U_{j}^{\prime}\right)  +\delta_{j2}\left(
U_{i}-U_{i}^{\prime}\right)  \right]  $, the trace of which is $4F\cos\left(
k_{f}r_{1}/2\right)  \cos\left(  k_{f}X_{1}\right)  \left(  U_{2}%
-U_{2}^{\prime}\right)  $. \ The $X$-space average of the first term in
$\Phi_{ij}$ is $2F\delta_{i2}\cos\left(  k_{f}r_{1}/2\right)  \frac{1}{V}%
\int\left[  \int\int\left(  u_{j}-u_{j}^{\prime}\right)  dX_{2}dX_{3}\right]
\cos\left(  k_{f}X_{1}\right)  dX_{1}$; interchange $i$ and $j$ to obtain the
second term in $\Phi_{ij}$. \ Whichever average is employed, this force
introduces a term that has no small-scale spatial variation and is negligible
in our scale analysis.

\qquad Forced turbulence is temporally intermittent such that a space average,
e.g., $\left\langle d_{ii}\right\rangle _{\mathbb{R}}$, does not obey
$\partial_{t}\left\langle d_{ii}\right\rangle _{\mathbb{R}}=0$. \ The temporal
intermittency observed by Borue and Orszag (1996) illustrates this fact; of
particular relevance is the observation of repeated events characterized by
accumulation of space-averaged energy in their mean flow (defined by a surface
average in their calculation), followed by a burst of transfer of energy from
their mean flow to the space-averaged turbulent energy. \ Given the conditions
mentioned below (\ref{tempavederiv}), one can time-average
(\ref{spacapproxlasttrac}) such that$\left\langle \partial_{t}\left\langle
\widehat{d}_{ii}\right\rangle _{\mathbb{R}}\right\rangle _{T}$ can be
neglected; the time average has the effect of averaging the temporal
intermittency. \ Now, apply to (\ref{spacapproxlasttrac}) the $r$-sphere
volume average (\ref{r-sphere}). \ Consider the case in which the Reynolds
number is large enough that $r_{S}$ is in the inertial range, then we can
neglect the term proportional to $\nu$ in (\ref{spacapproxlasttrac}), and as
shown in the preceding paragraph, any forcing term can be neglected. \ For a
sufficiently long time average we have the approximation that $\oint_{r_{n}%
}\left\langle \left\langle \widehat{d}_{iin}\right\rangle _{\mathbb{R}%
}\right\rangle _{T}\simeq-\frac{4}{3}\left\langle \left\langle \widehat
{\varepsilon}\right\rangle _{\mathbb{R}}\right\rangle _{T}r_{S}$ (recall that
this is based on neglecting the time-derivative and viscous terms in
(\ref{spacapproxlasttrac}) and the forcing because the forcing has no
small-scale spatial variation and is therefore negligible in our scale
analysis). \ A similar generalization of Kolmogorov's 4/5 law, namely,
$\oint_{r_{n}}D_{iin}\simeq-\frac{4}{3}\left\langle \varepsilon\right\rangle
_{E}r$, was obtained in Lindborg (1996), Frisch (1995), and Hill (1997) for
the inertial range of homogeneous, anisotropic turbulence.

\bigskip

\begin{center}
\textbf{Appendix B:\ Data}\bigskip

\textbf{B.1 Inner Scales}\bigskip
\end{center}

\qquad Inner scale was first defined by Obukhov (1949) as the $r$ at which the
asymptotic formulas for the inertial and viscous ranges are equal. \ Inner
scales are more applicable in our scaling analysis by a factor of about 10
compared with $\eta$. \ Inner scales for $D_{11}$ and $D_{\beta\beta}$,
denoted $\ell_{11}$ and $\ell_{\beta\beta}$, can be related to $\eta$ using
$\left\langle \varepsilon\right\rangle _{E}=15\nu\left\langle \left(
\partial_{1}u_{1}\right)  ^{2}\right\rangle _{E}=\left(  15/2\right)
\nu\left\langle \left(  \partial_{1}u_{\beta}\right)  ^{2}\right\rangle _{E}$,
which is valid on the basis of local isotropy and incompressibility. \ Then,
$\ell_{11}=(2/3)^{3/4}\ell_{\beta\beta}=13\eta$. \ Inner scales for $D_{111}$
and $D_{1\beta\beta}$, denoted $\ell_{111}$ and $\ell_{1\beta\beta}$, can be
related to $\eta$ on the additional empirical basis that the derivative
skewness, $\left\langle \left(  \partial_{1}u_{1}\right)  ^{3}\right\rangle
_{E}/\left\langle \left(  \partial_{1}u_{1}\right)  ^{2}\right\rangle
_{E}^{3/2}$, varies little from $-0.5$ over observed values of Reynolds number
(Sreenivasan and Antonia, 1997; Belin \textit{et al.}, 1997). \ Then
$\ell_{111}=\sqrt{2}\ell_{1\beta\beta}=9.6\eta$. \ The average of these four
inner scales is$\ \left(  \ell_{11}+\ell_{\beta\beta}+\ell_{111}+\ell
_{1\beta\beta}\right)  /4=10\eta$. \ Even though the turbulence being studied
need not be locally isotropic, we define $\ell\equiv10\eta$. \ This is the
background of definition (\ref{elscale}).

\bigskip

\begin{center}
\textbf{B.2 Typical Data}\bigskip
\end{center}

\qquad Data are needed for the investigations in Sec. 7.4. \ Let
$\mathbf{r}/r$, $\breve{\imath}$, and $\breve{e}$ be orthogonal unit vectors.
\ Let subscript $1$ denote projection in the direction $\mathbf{r}/r$, e.g.,
$D_{11}\equiv\left(  r_{i}/r\right)  \left(  r_{j}/r\right)  D_{ij}$. \ Let
subscript $\beta$ denote projection in either the $\breve{\imath}$ or
$\breve{e}$ directions (we need not distinguish which direction), e.g.,
$D_{\beta\beta1}$ is either $\breve{\imath}_{i}\breve{\imath}_{j}\left(
r_{k}/r\right)  D_{ijk}$ or $\breve{e}_{i}\breve{e}_{j}\left(  r_{k}/r\right)
D_{ijk}$, but not $\breve{e}_{i}\breve{\imath}_{j}\left(  r_{k}/r\right)
D_{ijk}$. \ If a distinction need not be made as to the direction of
projection, then subscript $\alpha$ is used; thus, $D_{\alpha\alpha}$ is
either $D_{11}$ or $D_{\beta\beta}$. \ No summation is implied by repeated
Greek indices. \ A unit vector $\breve{a}$ in an arbitrary direction is a
linear combination of the unit vectors $\mathbf{r}/r$, $\breve{\imath}$, and
$\breve{e}.$ \ Thus, if projections of a quantity have the same order of
magnitude and sign in all three directions $\mathbf{r}/r$, $\breve{\imath}$,
and $\breve{e}$, then the projection in an arbitrary direction $\breve{a}$
also has that order of magnitude.

\qquad For the inertial range we use the formulas
\begin{align}
D_{\alpha\alpha}  &  =\left\langle \varepsilon\right\rangle _{E}^{2/3}%
r^{2/3}K_{\alpha\alpha}\left(  \mathbf{X,r},t\right)  , \tag{B1}%
\label{inertial2}\\
D_{111}  &  =-\left\langle \varepsilon\right\rangle _{E}rK_{111}\left(
\mathbf{X,r},t\right)  \text{ , \ \ \ \ \ \ }D_{1\beta\beta}=-\left\langle
\varepsilon\right\rangle _{E}rK_{1\beta\beta}\left(  \mathbf{X,r},t\right)  .
\tag{B2}\label{inertial3}%
\end{align}
The dimensionless coefficient functions, $K_{\alpha\alpha}\left(
\mathbf{X,r},t\right)  $, $K_{111}\left(  \mathbf{X,r},t\right)  $, and
$K_{1\beta\beta}\left(  \mathbf{X,r},t\right)  $, are included to emphasize
that our inertial-range data, like real data, need not be precisely
homogeneous, locally isotropic, or stationary. \ The coefficient functions are
assumed to be of the order of unity, and when differentiating the structure
functions with respect to $r_{i},$ the derivatives of the coefficient
functions are assumed to be negligible compared to the derivative of $r^{2/3}$
in (\ref{inertial2}) and $r$ in (\ref{inertial3}).\ \ As motivation for this
assumption, consider that for the case of local isotropy the above coefficient
functions are constants between 2.7 and 0.26. \ The choice to scale with
$\ell\equiv10\eta,$ rather than with $\eta,$ causes the coefficient functions
to be of the order of unity.

\qquad The slight effect of intermittency on the exponent 2/3 in
(\ref{inertial2}) is not of\ concern here. Of more significance is the finding
by Mydlarski and Warhaft (1996) of power-law ranges that are precursors to the
inertial range. \ Their precursor power-law exponents are smaller than the 2/3
exponent of the inertial range, and the precursor exponents approach 2/3 as
Reynolds number increases. \ Our scale analysis can be extended to apply for
those weaker power laws; the accuracy of the scaling condition would be
correspondingly weakened.

\qquad Using $\ell$ defined in (\ref{elscale}), the viscous-range formulas for
the scale analysis are
\begin{align}
D_{\alpha\alpha}  &  =\left\langle \varepsilon\right\rangle _{E}^{2/3}%
\ell^{2/3}\left(  r/\ell\right)  ^{2}k_{\alpha\alpha}\left(  \mathbf{X,r}%
,t\right)  \text{, \ \ }.\tag{B3}\label{viscous}\\
D_{111}  &  =-\left\langle \varepsilon\right\rangle _{E}\ell\left(
r/\ell\right)  ^{3}k_{111}\left(  \mathbf{X,r},t\right)  \text{,
\ \ }D_{1\beta\beta}=-\left\langle \varepsilon\right\rangle _{E}\ell\left(
r/\ell\right)  ^{3}k_{1\beta\beta}\left(  \mathbf{X,r},t\right)  ,
\tag{B4}\label{viscous3}%
\end{align}
where the dimensionless coefficient functions, $k_{\alpha\alpha}\left(
\mathbf{X,r},t\right)  $, $k_{111}\left(  \mathbf{X,r},t\right)  $, and
$k_{1\beta\beta}\left(  \mathbf{X,r},t\right)  $ are assumed to be of the
order of unity, and when differentiating the structure functions with respect
to $r_{i}$ the derivatives of the coefficient functions are assumed to be
negligible. \ In support of this assumption, note that for the case of local
isotropy these coefficient functions are constants between 2.9 and 0.57. \ The
choice to scale with $\ell\equiv10\eta,$ rather than with $\eta,$ causes the
coefficient functions to be of the order of unity.

\qquad For $r$ between the inertial and viscous ranges, the structure
functions $D_{\alpha\alpha}$, $D_{111}$, and $D_{1\beta\beta}$ have monotonic
transitions between the asymptotic formulas (\ref{inertial2}),
(\ref{inertial3}), and (\ref{viscous}), (\ref{viscous3}). \ Therefore, if a
quantity is negligible on the basis of both (\ref{inertial2}),
(\ref{inertial3}), and (\ref{viscous}), (\ref{viscous3}), then it is
negligible for all $r$ from within the inertial range to within the viscous range.

\qquad Consider the projection of $\partial_{r_{n}}D_{ijn}$\ in directions
parallel to $\mathbf{r}$, i.e., $\left(  r_{i}/r\right)  \left(
r_{j}/r\right)  \partial_{r_{n}}D_{ijn},$\ and perpendicular to $\mathbf{r}$,
e.g., $\breve{e}_{i}\breve{e}_{j}\partial_{r_{n}}D_{ijn}$. \ Note that neither
projection commutes\ with the derivatives $\partial_{r_{i}}$, e.g., $\breve
{e}_{i}\breve{e}_{j}\left(  \partial_{r_{n}}D_{ijn}\right)  \neq
\partial_{r_{n}}\left(  \breve{e}_{i}\breve{e}_{j}D_{ijn}\right)  $. \ For
example, if $D_{ijn}$ is locally isotropic, then differentiating the
isotropic-tensor formula gives $\breve{e}_{i}\breve{e}_{j}\left(
\partial_{r_{n}}D_{ijn}\right)  =\partial_{r}D_{\beta\beta1}+\frac{4}%
{r}D_{\beta\beta1}$, whereas $\partial_{r_{n}}\left(  \breve{e}_{i}\breve
{e}_{j}D_{ijn}\right)  =\partial_{r_{n}}D_{\beta\beta n}+\frac{2}{r}%
D_{\beta\beta1}$; another example is: $\left(  r_{i}/r\right)  \left(
r_{j}/r\right)  \left(  \partial_{r_{n}}D_{ijn}\right)  =\partial_{r}%
D_{111}+\frac{2}{r}D_{111}-\frac{4}{r}D_{\beta\beta1},$ whereas $\partial
_{r_{n}}\left[  \left(  r_{i}/r\right)  \left(  r_{j}/r\right)  D_{ijn}%
\right]  =\partial_{r_{n}}D_{11n}=\partial_{r}D_{111}+\frac{2}{r}D_{111}$.
\ By use of (\ref{inertial3}), both $\breve{e}_{i}\breve{e}_{j}\left(
\partial_{r_{n}}D_{ijn}\right)  $ and $\left(  r_{i}/r\right)  \left(
r_{j}/r\right)  \partial_{r_{n}}D_{ijn}$ are $-\frac{4}{3}\left\langle
\varepsilon\right\rangle _{E}$ for the locally isotropic case. \ Since the
projections in all three directions are the same, the projection in the
arbitrary direction $\breve{a}$, i.e., $\breve{a}_{i}\breve{a}_{j}%
\partial_{r_{n}}D_{ijn}$, also equals $-\frac{4}{3}\left\langle \varepsilon
\right\rangle _{E}$ for the locally isotropic case. \ Although our data are
not locally isotropic, assume that for our data $\breve{a}_{i}\breve{a}%
_{j}\partial_{r_{n}}D_{ijn}$ is of the order of $-\left\langle \varepsilon
\right\rangle _{E}$ in the inertial range. \ For the viscous range and for the
locally isotropic case, $\breve{e}_{i}\breve{e}_{j}\partial_{r_{n}}%
D_{ijn}=\partial_{r}D_{\beta\beta1}+\frac{4}{r}D_{\beta\beta1}=\frac{7}%
{r}D_{\beta\beta1}=\frac{14}{3r}D_{111}$ and $\left(  r_{i}/r\right)  \left(
r_{j}/r\right)  \partial_{r_{n}}D_{ijn}=\partial_{r}D_{111}+\frac{2}{r}%
D_{111}-\frac{4}{r}D_{\beta\beta1}=\frac{7}{3r}D_{111}$; these formulas
combined with (\ref{viscous3}) imply that $\breve{a}_{i}\breve{a}_{j}%
\partial_{r_{n}}D_{ijn}$ is of the order of $-\left\langle \varepsilon
\right\rangle _{E}\left(  r/\ell\right)  ^{2}$ for the viscous range. \ Assume
that this is true of our data as well.

\qquad The same manipulations apply to the projection of $\partial_{r_{n}%
}\partial_{r_{n}}D_{ij}$. \ Differentiation of the isotropic-tensor formulas
gives $\left(  r_{i}/r\right)  \left(  r_{j}/r\right)  \partial_{r_{n}%
}\partial_{r_{n}}D_{ij}=\left(  \partial_{r}+\frac{2}{r}\right)  \partial
_{r}D_{11}+\frac{4}{r^{2}}\left(  D_{\beta\beta}-D_{11}\right)  $ [in
contrast, $\partial_{r_{n}}\partial_{r_{n}}D_{11}=\left(  \partial_{r}%
+\frac{2}{r}\right)  \partial_{r}D_{11}$], and $\breve{e}_{i}\breve{e}%
_{j}\partial_{r_{n}}\partial_{r_{n}}D_{ij}=\left(  \partial_{r}+\frac{2}%
{r}\right)  \partial_{r}D_{\beta\beta}-\frac{2}{r^{2}}\left(  D_{\beta\beta
}-D_{11}\right)  $. \ By use of (\ref{inertial2}) and (\ref{viscous}) we find
for the projection in an arbitrary direction $\breve{a}$, that $\breve{a}%
_{i}\breve{a}_{j}\partial_{r_{n}}\partial_{r_{n}}D_{ij}$ is about
$2\left\langle \varepsilon\right\rangle _{E}^{2/3}r^{-4/3}$ in an inertial
range and is about $10\left\langle \varepsilon\right\rangle _{E}^{2/3}%
\ell^{-4/3}\simeq\left\langle \varepsilon\right\rangle _{E}/2\nu$ in the
viscous range. \ Assume that this is true of our data as well.

\qquad In effect, our definition of the inertial range includes
(\ref{inertial2}), (\ref{inertial3}), and that the projections in an arbitrary
direction $\breve{a}$ of the two terms in (\ref{exactave}) behave as \
\begin{align}
\breve{a}_{i}\breve{a}_{j}\partial_{r_{n}}D_{ijn}  &  \sim-\left\langle
\varepsilon\right\rangle _{E},\tag{B5}\label{arbproD3div}\\
2\nu\breve{a}_{i}\breve{a}_{j}\partial_{r_{n}}\partial_{r_{n}}D_{ij}  &
\sim0.2\left\langle \varepsilon\right\rangle _{E}\left(  r/\ell\right)
^{-4/3}, \tag{B6}\label{arbproD2lapinert}%
\end{align}
where $\sim$ means ``is of the order of.'' \ Our definition of the viscous
range includes (\ref{viscous}) and (\ref{viscous3}), and that the projections
in an arbitrary direction behave as
\begin{align}
\breve{a}_{i}\breve{a}_{j}\partial_{r_{n}}D_{ijn}  &  \sim-\left\langle
\varepsilon\right\rangle _{E}\left(  r/\ell\right)  ^{2}, \tag{B7}%
\label{arbproD3divvisc}\\
2\nu\breve{a}_{i}\breve{a}_{j}\partial_{r_{n}}\partial_{r_{n}}D_{ij}  &
\sim\left\langle \varepsilon\right\rangle _{E}. \tag{B8}%
\label{arbproD2lapvisc}%
\end{align}
\ For both ranges we include the additional assumption that the off-diagonal
components $D_{\alpha\beta}$ (for $\alpha\neq\beta$) are not greater in
magnitude than $D_{\alpha\alpha}$.

\qquad We assume that diagonal components of $E_{ij}\left(  \mathbf{X,r}%
,t\right)  $ are of order $\left\langle \varepsilon\right\rangle _{E}$; that
is, when projected on a arbitrary direction $\breve{a}$,
\begin{equation}
\breve{a}_{i}\breve{a}_{j}E_{ij}\sim\left\langle \varepsilon\right\rangle
_{E}. \tag{B9}\label{ediagonal}%
\end{equation}
\ In support of this assumption, recall that $E_{ii}=4\left\langle
\varepsilon\right\rangle _{E}$ on the basis of homogeneity, and, in the case
of local isotropy, $E_{\alpha\alpha}\equiv4\left\langle \varepsilon
\right\rangle _{E}/3$.

\qquad Data are needed for the diagonal elements $\breve{a}_{i}\breve{a}%
_{j}T_{ij}$. \ Because $\left\langle \left(  p-p^{\prime}\right)  \left(
u_{j}-u_{j}^{\prime}\right)  \right\rangle _{E}$ vanishes on the basis of
local isotropy and because $\partial_{X_{in}}$ operating on any average
vanishes on the basis of homogeneity, it is assumed that
\begin{equation}
\left|  \breve{a}_{i}\breve{a}_{j}\left\langle 2\left(  p-p^{\prime}\right)
\left(  s_{ij}-s_{ij}^{\prime}\right)  \right\rangle _{E}\right|  \geq\left|
\breve{a}_{i}\breve{a}_{j}\partial_{X_{i}}\left\langle \left(  p-p^{\prime
}\right)  \left(  u_{j}-u_{j}^{\prime}\right)  \right\rangle _{E}\right|  .
\tag{B10}\label{Tterm}%
\end{equation}
Of course, this is not true for the sum of diagonal components because of
(\ref{taucont}). \ It is likely that the right side of (\ref{Tterm}) is much
smaller than the left side, but a more restrictive condition than
(\ref{Tterm}) is not needed. \ On the basis of DNS, Borue and Orszag (1996)
show the cross spectrum of velocity and pressure gradient, where both velocity
and pressure gradient are projected in their $y$-direction. \ For the inertial
range, their data show that the corresponding diagonal component of $T_{ij}$
is proportional to $\left\langle \varepsilon\right\rangle _{E}r/L$. \ More
details for other flows would be welcome because $T_{ij}$ vanishes for locally
isotropic turbulence (Hill, 1997). \ Therefore, its anisotropic behavior is of
interest. \ However, based on the result by Borue and Orszag (1996), it is
assumed that our data in the inertial range obeys
\begin{equation}
\breve{a}_{i}\breve{a}_{j}T_{ij}\sim\left\langle \varepsilon\right\rangle
_{E}\left(  r/L\right)  . \tag{B11}\label{Tinertial1}%
\end{equation}
The data by Alvelius and Johansson (2000) are consistent with
(\ref{Tinertial1}). \ Using data from nearly homogeneous turbulence, Lindborg
(1996) found that the single-point pressure strain correlation has a
longitudinal component that is approximately $-4\left\langle \varepsilon
\right\rangle _{E}/3$\ and a transverse component that is approximately
$2\left\langle \varepsilon\right\rangle _{E}/3$. For homogeneous turbulence in
the limit $r\rightarrow\infty$\ Lindborg's result corresponds to
$T_{11}\rightarrow-16\left\langle \varepsilon\right\rangle _{E}/3$ and
$T_{\beta\beta}\rightarrow-8\left\langle \varepsilon\right\rangle _{E}/3$;
this agrees in order of magnitude with (\ref{Tinertial1}) evaluated at $r=L $.
\ The first nonvanishing term of the Taylor series expansion of $\breve{a}%
_{i}\breve{a}_{j}\left\langle 2\left(  p-p^{\prime}\right)  \left(
s_{ij}-s_{ij}^{\prime}\right)  \right\rangle _{E}$ is $r^{2}$ times the
average of the product of pressure gradient and strain-rate gradient. \ This
suggests that for the viscous range,
\begin{equation}
\breve{a}_{i}\breve{a}_{j}T_{ij}\sim\left\langle \varepsilon\right\rangle
_{E}\left(  r/\ell\right)  ^{2}\left(  \ell/L\right)  . \tag{B12}%
\label{Tviscous}%
\end{equation}
The form of (\ref{Tviscous}) is chosen to equal (\ref{Tinertial1})\ at
$r=\ell$. \ In the absence of further information, (\ref{Tviscous}) is assumed
to be valid.

\qquad Finally, data are needed for $\widehat{\Gamma}_{in}$ and $\widehat
{\Gamma}_{ijn},$ which are defined in (\ref{gammadef}) and appear in the
second term of (\ref{reyn}). \ As described in Sec. 7.4, the second term in
(\ref{reyn}) is important for the case of large-scale structures in the flow.
\ Assume that the experimenter chooses a flow that mitigates against
large-scale structure; grid-generated turbulence is an example. \ In this
case, it is assumed that
\begin{equation}
\widehat{\Gamma}_{in}\leq\upsilon^{2},\text{ and }\widehat{\Gamma}_{ijn}%
\ll\upsilon^{3}, \tag{B13}\label{gammapprox}%
\end{equation}
where $\upsilon$ is defined in (\ref{velscale}).

\qquad Relationships (\ref{inertial2})-(\ref{gammapprox}) serve as exemplary
data in the scale analysis. \ The fact that we have data only for projections
in an arbitrary direction $\breve{a}$ means that we can investigate only the
diagonal components of (\ref{exactave}). \ The off-diagonal components, which
are obtained by projection in two orthogonal directions, cannot be studied
here. \ Relationships like (\ref{inertial2})-(\ref{ediagonal}) are most often
associated with the assumption of local isotropy.\ \ However, like anemometry
data, these relationships can be fulfilled for coefficient functions [as
defined in (\ref{inertial2})-(\ref{viscous3})] of the order of unity without
the specific restrictions of local isotropy being precisely fulfilled. \ For
instance, for $D_{ij}$, the restrictions for local isotropy are that its
off-diagonal elements are zero, and that $\breve{\imath}_{i}\breve{\imath}%
_{j}D_{ij}=\breve{e}_{i}\breve{e}_{j}D_{ij}$, and that $D_{11}$ is related to
$D_{\beta\beta}$ by an incompressibility condition. \ In the scale analysis,
such restrictions are not used; therefore, local isotropy is not assumed.

\bigskip

\begin{center}
\bigskip\textbf{Appendix C:\ Homogeneity Implemented Using the Calculus of
Local Homogeneity}
\end{center}

\qquad Although homogeneity is mentioned only briefly in this study, it is
useful to introduce it and to show how the calculus of local homogeneity
produces the predictions of homogeneity for the case of homogeneous
turbulence. \ Homogeneity is the approximation that ensemble averages do not
depend on the position at which the average is obtained (Monin and Yaglom,
1975). \ That position being $\mathbf{X}$, we implement this approximation by
neglecting the result of $\partial_{X_{n}}$ operating on any average. \ For
example, in (\ref{exactave}) $\frac{1}{4}\partial_{X_{n}}\partial_{X_{n}%
}D_{ij}=0$, $\partial_{X_{n}}F_{ijn}=0$, and for the average of (\ref{tau2})
$\partial_{X_{i}}\left\langle \left(  p-p^{\prime}\right)  \left(  u_{j}%
-u_{j}^{\prime}\right)  \right\rangle _{E}=0$ such that $T_{ij}=-2\left\langle
\left(  p-p^{\prime}\right)  \left(  s_{ij}-s_{ij}^{\prime}\right)
\right\rangle _{E}$, from which we obtain $T_{ii}=0$ because $s_{ii}=0$ by
incompressibility. \ The Taylor series of $p(\mathbf{x},t\mathbf{)}$ around
point $\mathbf{X}$ is $p(\mathbf{x},t\mathbf{)}=p(\mathbf{X},t\mathbf{)+}%
(x_{n}-X_{n})\partial_{X_{n}}p(\mathbf{X},t\mathbf{)+\cdots}$. \ Upon
averaging, homogeneity requires that $\partial_{X_{n}}\left\langle
p(\mathbf{X},t\mathbf{)}\right\rangle _{E}=0$, etc., such that $\left\langle
p(\mathbf{x},t\mathbf{)}\right\rangle _{E}\mathbf{=}\left\langle
p(\mathbf{X},t\mathbf{)}\right\rangle _{E}$, and similarly $\left\langle
p(\mathbf{x}^{\prime},t\mathbf{)}\right\rangle _{E}\mathbf{=}\left\langle
p(\mathbf{X},t\mathbf{)}\right\rangle _{E}$; similarly, $\left\langle
\varepsilon\right\rangle _{E}=\left\langle \varepsilon^{\prime}\right\rangle
_{E}$. \ Within the average of (\ref{eii}) we have $\partial_{X_{n}}%
\partial_{X_{n}}\left\langle p+p^{\prime}\right\rangle _{E}=2\partial_{X_{n}%
}\partial_{X_{n}}\left\langle p(\mathbf{X},t\mathbf{)}\right\rangle _{E}=0$,
etc., such that $E_{ii}=4\left\langle \varepsilon\right\rangle _{E}$. \ From
(\ref{avesecdinc1}) homogeneity gives the incompressibility condition,
$\partial_{r_{n}}D_{jn}=0$.

\qquad In\textbf{\ }Sec. 3.3 the spatial average is a volume average in
$\mathbf{X}$-space such that the equations [e.g., (\ref{spatialaveext2}) and
(\ref{spatialavetrace})] do not contain $\partial_{X_{n}}$operating on an
average. \ For those spatially averaged equations, homogeneity can be
implemented by neglecting any average over the surface bounding the averaging
volume of the surface-normal component of any vector. \ The basis for this
implementation is that there are no net\ average fluxes in homogeneous turbulence.
\end{document}